\documentclass[showpacs,twocolumn,sort&compress,numbers,longbibliography]{revtex4-1}
\pdfoutput=1
\usepackage{bm}
\usepackage{amsmath}
\usepackage{amssymb}
\usepackage{graphicx}
\usepackage{slashed}
\usepackage{accents}
\usepackage{mathrsfs}
\usepackage{slashed}
\usepackage{braket}
\usepackage{units}
\usepackage{hyperref}
\usepackage{calc}	

\usepackage{slashed}
\usepackage{upgreek}

\makeatletter
\let\Re\@undefined
\let\Im\@undefined
\makeatother

\DeclareSymbolFont{eulerscript}{U}{eur}{m}{n}
\DeclareSymbolFontAlphabet{\matheuler}{eulerscript}
\newcommand{\T}{T}
\newcommand{\Q}{\mathcal{Q}}
\newcommand{\I}{i}
\newcommand{\nfrac}[2]{{#1}/{#2}}
\newcommand{\one}{\mathbf{1}}
\newcommand{\del}{\partial}
\newcommand{\eps}{\epsilon}
\newcommand{\mc}[1]{\mathcal{#1}}
\newcommand{\lb}{\left(}
\newcommand{\rb}{\right)}
\newcommand{\rsb}{\right]}
\newcommand{\lsb}{\left[}
\newcommand{\rcb}{\right\}}
\newcommand{\lcb}{\left\{}
\newcommand{\abs}[1]{\left|#1\right|}
\DeclareMathOperator{\tr}{\mathbf{tr}}
\DeclareMathOperator{\Re}{Re}
\DeclareMathOperator{\Im}{Im}
\DeclareMathOperator{\Ai}{Ai}
\DeclareMathOperator{\Gi}{Gi}
\DeclareMathOperator{\sinc}{sinc}
\newcommand{\PMNS}{\mathbf{U}}
\newcommand{\lplus}{{{}+}}
\newcommand{\lminus}{{{}-}}
\newcommand{\lone}{{\scalebox{.64}{$\matheuler{I}$}}}
\newcommand{\ltwo}{{\scalebox{.64}{$\matheuler{II}$}}}
\newcommand{\lperp}{\perp}
\newcommand{\s}[1]{\slashed{#1}}
\newcommand{\Ftilde}{\mathfrak{F}}

\newcommand{\sdist}{\kern 0.20em}

\newcommand{\secref}[1]{Sec.\sdist\ref{#1}}
\newcommand{\appref}[1]{App.\sdist\ref{#1}}
\renewcommand{\eqref}[1]{Eq.\sdist(\ref{#1})}
\newcommand{\figref}[1]{Fig.\sdist\ref{#1}}

\begin{document}
\title{Nonlinear neutrino-photon interactions inside strong laser pulses}
\author{Sebastian \surname{Meuren}}
\email{s.meuren@mpi-hd.mpg.de}
\author{Christoph H. \surname{Keitel}}
\email{keitel@mpi-hd.mpg.de}
\author{Antonino \surname{Di Piazza}}
\email{dipiazza@mpi-hd.mpg.de}
\affiliation{Max-Planck-Institut f\"ur Kernphysik, Saupfercheckweg 1, D-69117 Heidelberg, Germany}
\date{\today}

\begin{abstract}
Even though neutrinos are neutral particles and interact only via the exchange of weak gauge bosons, charged leptons and quarks can mediate a coupling to the photon field beyond tree level. Inside a relativistically strong laser field nonlinear effects in the laser amplitude can play an important role, as electrons and positrons interact nonperturbatively with the coherent part of the photon field. Here, we calculate for the first time the leading-order contribution to the axial-vector--vector current-coupling tensor inside an arbitrary plane-wave laser field (which is taken into account exactly by employing the Furry picture). The current-coupling tensor appears in the calculation of various electroweak processes inside strong laser fields like photon emission or trident electron-positron pair production by a neutrino. Moreover, as we will see below, the axial-vector--vector current-coupling tensor contains the Adler-Bell-Jackiw (ABJ) anomaly. This occurrence renders the current-coupling tensor also interesting from a fundamental point of view, as it is  the simplest Feynman diagram in an external field featuring this kind of anomaly. 
\end{abstract}

\maketitle
\section{Introduction}

As different neutrino mass eigenstates exist \cite{olive_review_2014,bilenky_introduction_2010,avignone_double_2008,giunti_fundamentals_2007,mohapatra_massive_2004,hohler_neutrino_2003,bilenky_massive_1987}, only the lowest one is stable and all others can, in principle, decay radiatively \cite{bilenky_lepton_1977,lee_natural_1977,marciano_exotic_1977,petcov_processes_1977,shrock_electromagnetic_1982,pal_radiative_1982}. However, due to the smallness of the available phase space and the Glashow-Iliopoulos-Maiani (GIM) suppression mechanism (i.e. cancellations between contributions from different fermion generations) \cite{glashow_weak_1970,pal_radiative_1982} the neutrino life time is much larger than the age of the universe (the electromagnetic properties of neutrinos in vacuum are discussed in \cite{nieves_electromagnetic_1982,robert_electromagnetic_1982,dvornikov_electromagnetic_2004,dvornikov_electric_2004,giunti_neutrino_2009,broggini_electromagnetic_2012}). 

Nevertheless, a neutrino can emit photons inside strong electromagnetic background fields, which catalyze the decay. For example, strong magnetic fields encountered in various astrophysical situations substantially reduce the neutrino life time \cite{gvozdev_magnetic_1992,zhukovsky_radiative_1996,kachelries_radiative_1997,gvozdev_resonance_1997,ioannisian_cherenkov_1997,anikin_radiative_2013,ternov_neutrino_2014} (see also \cite{gvozdev_radiative_1994}, where the Coulomb field has been investigated). Inside background fields also the production of electron-positron pairs -- which is not possible in vacuum due to energy-momentum conservation -- is feasible under certain circumstances \cite{choban_production_1969,borisov_electron-positron_1993,kuznetsov_neutrino_1997,kuznetsov_lepton_2000,kuznetzov_lepton-pair_2002,tinsley_pair_2005,dicus_pair_2007}.

Moreover, neutrino properties like their mass and their magnetic moment are modified by electromagnetic background fields \cite{kuznetsov_neutrino_2006,erdas_neutrino_2009,dobrynina_neutrino_2013}. Implications for neutrino oscillations have been studied in \cite{egorov_neutrino_2000,dvornikov_neutrino_2001,lobanov_neutrino_2001,dvornikov_parametric_2004} and the possibility for spin light has been pointed out in \cite{lobanov_spin_2003,studenikin_neutrino_2005}. 

The presence of electromagnetic background fields could also be exploited to create neutrinos, e.g. via photon splitting \cite{skobelev_the_1972,deraad_photon_1976,kuznetsov_photon_1998}, scattering \cite{shaisultanov_photon-neutrino_1998,chistyakov_photon-neutrino_2002,kuznetsov_process_2003} or the trident process \cite{titov_neutrino_2011} (for a review of electroweak processes in electromagnetic background fields see \cite{kuznetsov_electroweak_2013,kuznetsov_electroweak_2003}).

\begin{figure}[b!]
\centering
\begin{minipage}[c]{0.45\columnwidth}
\centering
\includegraphics{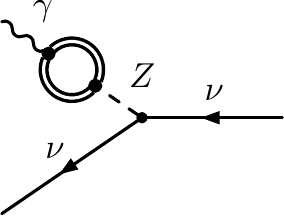}
\end{minipage}
\hspace*{-10pt}
\begin{minipage}[c]{0.45\columnwidth}
\centering
\includegraphics{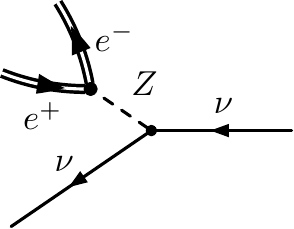}
\end{minipage}
\\\vspace*{10pt}
\begin{minipage}[c]{0.45\columnwidth}
\centering
\small\textbf{a}) Photon emission
\end{minipage}
\hspace*{-10pt}
\begin{minipage}[c]{0.45\columnwidth}
\centering
\small\textbf{b}) Pair production
\end{minipage}
\\\vspace*{10pt}
\begin{minipage}[c]{0.7\columnwidth}
\centering
\includegraphics{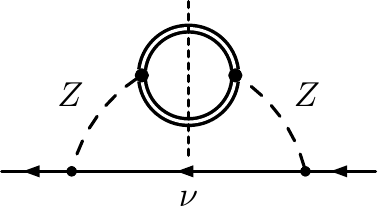}
\\
\small\textbf{c}) Optical theorem for pair production
\end{minipage}
\caption{\label{fig:neutrinoprocesses}\textbf{a}) Photon emission and \textbf{b}) trident electron-positron pair production by a neutrino inside a strong, plane-wave laser field (mediated by the neutral current, i.e. $Z$ boson exchange). For an electron neutrino also the charged current must be taken into account ($W$ boson exchange, see \figref{fig:neutrinophotonvertex}). \textbf{c}) The total trident pair-production probability is related to the imaginary part of the neutrino self-energy diagram (see e.g. \cite{meuren_polarization-operator_2015,borisov_electron-positron_1993} for details). The double lines denote here electron and positron states, which are dressed by the laser field (time axis from right to left).}
\end{figure}

It is an interesting question whether the emission of photons by neutrinos or other processes like electron-positron pair production could be investigated in a laboratory experiment using high-power lasers (see \figref{fig:neutrinoprocesses}). To shed light on the feasibility of this idea, the special case of a circularly polarized, monochromatic plane-wave laser field has been analyzed in \cite{gvozdev_radiative_1993,gvozdev_electromagnetic_1994,gvozdev_radiative_1996} (see also \cite{skobelev_interaction_1991}). As lasers field are naturally produced with linear polarization and the hightest intensities can only be achieved by using short laser pulses, it is desirable to generalize these results accordingly. In the present paper we will therefore consider a plane-wave laser field with arbitrary polarization and pulse shape.

Inside plane-wave laser fields the probability for a neutrino process depends primarily on the laser intensity and the neutrino energy. A convenient gauge- and Lorentz-invariant measure for the laser field strength is given by the parameter $\xi=\nfrac{|e|E_0}{(m\omega c)}$, where $E_0$ is the electric field amplitude and $\omega$ the central angular frequency of the laser ($e<0$ and $m$ denote the electron charge and mass, respectively). In the regime $\xi\gtrsim 1$ the interaction between the background field and the electron and the positron, must be taken into account exactly by solving the Dirac equation in the presence of the background field \cite{di_piazza_extremely_2012,dittrich_probingquantum_2000,fradkin_quantum_1991,ritus_1985,mitter_quantum_1975}. For a plane-wave field this is possible analytically and one obtains the Volkov states as single-particle states \cite{volkov_ueber_1935,landau_quantum_1981}. Working in momentum space, the only necessary modification of the Feynman rules is the replacement of the free vertex by the so-called dressed vertex (i.e. $-\I e\gamma^\mu \to \Gamma^\mu$ for QED; see \appref{sec:dressedvertexappendix} and e.g. \cite{meuren_polarization_2013,mitter_quantum_1975} for more details). Unlike in vacuum, four-momentum is conserved only up to a multiple of the laser four-momentum at the dressed vertex, which changes the kinematics of the processes.

It is well known that for $\xi \gg 1$ the formation region for single-vertex processes primed by the laser field is much smaller than the laser wavelength, such that the local constant-crossed field approximation is applicable \cite{di_piazza_extremely_2012,ritus_1985}. Therefore, the case of a constant-crossed background field (studied e.g. in \cite{galtsov_photoneutrino_1972,borisov_electron-positron_1993,gvozdev_radiative_1996,borovkov_oneloop_1999}) is particularly interesting and provides the order of magnitude for the expected probabilities. Inside a constant-crossed field the probability depends nontrivially only on the quantum-nonlinearity parameter $\chi = (2\nfrac{E_\nu}{mc^2}) (\nfrac{E_0}{E_{\mathrm{cr}}})$, where $E_\nu$ denotes the energy of the incoming neutrino and $E_{\mathrm{cr}}=\nfrac{m^2c^3}{(\hbar|e|)} = \unitfrac[1.3\times 10^{16}]{V}{cm}$ the critical field strength of QED \cite{schwinger_gauge_1951,heisenberg_folgerungen_1936,sauter_uber_1931} (the expression of $\chi$ given here assumes a head-on collision and neglects the neutrino mass). 

As the nonlinear-quantum parameter is inversely proportional to the cube of the electron (positron) mass ($\chi \sim m^{-3}$), nonlinear quantum effects caused by muon or tau leptons are strongly suppressed for reasonable parameters and ignored here. Correspondingly, the symmetry between different lepton generations is broken and the GIM mechanism does not apply. Furthermore, the laser provides additional energy and momentum to the reaction, which enlarges the available phase space. Due to these two reasons the probability for photon emission by neutrinos inside a plane-wave field is strongly enhanced in comparison with the vacuum case (note that the laser field also affects tree-level processes like the decay of a muon \cite{dicus_muon_2009,farzinnia_muon_2009}). Nevertheless, since the enhancement is primed by an electromagnetic exchange of photons between an electron/positron loop and the laser, we expect that the probabilities for nonlinear neutrino processes inside laser fields still contain the suppression factor $(\nfrac{m}{M_{Z,W}})^4 \sim 10^{-20}$ and an experimental observation is challenging ($M_Z \approx \unit[91]{GeV}$ and $M_W \approx \unit[80]{GeV}$ denote the mass of the $Z$ and the $W$ boson, respectively \cite{olive_review_2014}). 

By combining accelerator-based neutrino beams with energies in the GeV range \cite{ISSneutrino,IDSneutrino,MAP,kaplan_muon_2014,bogomilov_neutrino_2014,geer_muon_2012,group_accelerator_2009,bandyopadhyay_physics_2009,berg_cost-effective_2006,geer_neutrino_1998} with strong optical lasers ($\xi \sim 10^{2-3}$) \cite{ELI,CLF,XCELS,yanovsky_ultra_2008}, the nonlinear quantum regime $\chi\gtrsim 1$ could be entered, where also the production of real electron-positron pairs via the trident process becomes feasible \cite{choban_production_1969,borisov_electron-positron_1993,kuznetsov_neutrino_1997,kuznetsov_lepton_2000,kuznetzov_lepton-pair_2002,tinsley_pair_2005,dicus_pair_2007}. As the energy and momentum required to bring the electron-positron pair on shell are provided by the laser field, the probability for trident pair production even exceeds the one for photon emission if $\chi\gtrsim 1$ (the corresponding Feynman diagram contains only two interaction vertices, see \figref{fig:neutrinoprocesses}).

\begin{figure}
\centering
\begin{minipage}[t]{1.0\columnwidth}
\centering
\includegraphics{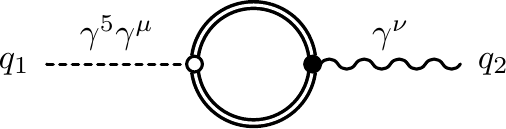}
\\\vspace*{4pt}
\small\textbf{a}) Current-coupling tensor $T_5^{\mu\nu}(q_1,q_2)$
\end{minipage}
\\\vspace*{15pt}
\begin{minipage}[t]{1.0\columnwidth}
\centering
\includegraphics{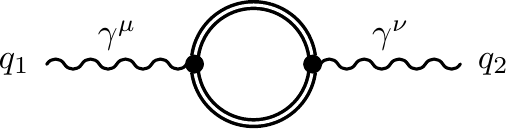}
\\\vspace*{4pt}
\small\textbf{b}) Polarization tensor $T^{\mu\nu}(q_1,q_2)$
\end{minipage}
\caption{\label{fig:T5vsT}\textbf{a}) The leading-order Feynman diagram for the coupling between the vector current ($\gamma^\mu$-vertex) and the axial-vector current ($\gamma^5\gamma^\mu$-vertex), see \eqref{eqn:axialvectorcouplingtensor}. \textbf{b}) The current-coupling tensor $T_5^{\mu\nu}(q_1,q_2)$ is closely related to the polarization tensor $T^{\mu\nu}(q_1,q_2)$, which was considered e.g. in \cite{gies_laser_2014,dinu_photon_2014,dinu_vacuum_2014,meuren_polarization_2013,baier_interaction_1975,becker_vacuum_1975}, see \eqref{eqn:polarizationoperator}. Solid lines indicate fermions, double lines Volkov states (which take the plane-wave background field exactly into account), wiggly lines photons and dashed lines the axial-vector current.}
\end{figure}

\begin{figure}
\centering
\begin{minipage}[t]{1.0\columnwidth}
\centering
\includegraphics{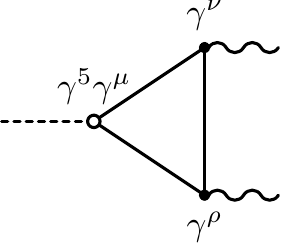}
\\\vspace*{4pt}
\textbf{a}) Anomalous triangle diagram
\end{minipage}
\\\vspace*{15pt}
\begin{minipage}[t]{1.0\columnwidth}
\centering
\includegraphics{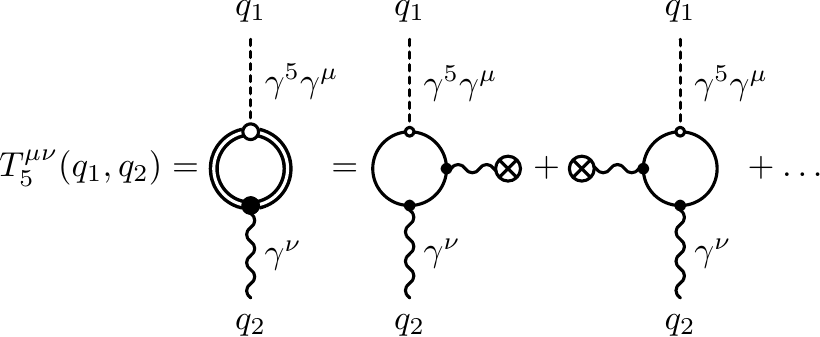}
\\\vspace*{4pt}
\textbf{b}) Leading-order expansion in the external field
\end{minipage}
\caption{\label{fig:anomaly}\textbf{a}) The axial-vector anomaly in vacuum QED is caused by the triangle diagram. \textbf{b}) As for weak external fields $A^\mu(\phi)$ (denoted by $\otimes$) the leading-order field-dependent contribution to the current-coupling tensor $T^{\mu\nu}_5(q_1,q_2)$ [see \eqref{eqn:axialvectorcouplingtensor}] corresponds to the triangle diagram, one also expects an anomalous term in the Ward-Takahashi identity for $T^{\mu\nu}_5(q_1,q_2)$ [see \eqref{eqn:T5wardidentity}]. Here, solid lines indicate the vacuum states and double lines dressed Volkov states for the charged fermions, wiggly lines photons and dashed lines the axial-vector current.}
\end{figure}

In order to calculate the probability for neutrino photon emission or trident pair production (via the optical theorem), the coupling between the vector current ($\gamma^\mu$-vertex) and the axial-vector current ($\gamma^5\gamma^\mu$-vertex) described by the tensor $T_5^{\mu\nu}(q_1,q_2)$ must be determined (see \figref{fig:T5vsT}). For a constant background field it has been investigated in \cite{gies_axial_2000,gies_neutrino_2000,shaisultanov_2000,schubert_vacuum_2000,schubert_vacuum_2000-1,gies_vacuum_2001,bhattacharya_2003,nieves_electromagnetic_2003}. In the present paper an arbitrary plane-wave laser field is considered as background field (see \secref{sec:planewavefields}) and a triple-integral representation for $T_5^{\mu\nu}(q_1,q_2)$ is derived, which can be transformed into a double-integral representation using the relations given in \cite{meuren_polarization-operator_2015}. Special attention is payed to the Ward-Takahashi identity, which contains a contribution due to the Adler-Bell-Jackiw (ABJ) anomaly (see \figref{fig:anomaly}) \cite{adler_axial-vector_1969,bell_pcac_1969}. The anomalous term is calculated explicitly by applying a suitable regularization procedure. 

The present paper is organized as follows: In \secref{sec:neutrinoradiation} the interaction between neutrinos and photons inside a plane-wave background field is considered and it is shown how the axial-vector--vector current-coupling tensor $T_5^{\mu\nu}(q_1,q_2)$ (see \figref{fig:T5vsT}) appears naturally in the electroweak sector of the standard model if plane-wave background fields are taken into account. The calculation of $T_5^{\mu\nu}(q_1,q_2)$ is then presented in \secref{sec:T5}, followed by a detailed discussion of the ABJ anomaly in \secref{sec:anomaly}. Subsequently, various important special cases like a constant-crossed and a circularly polarized, monochromatic field are considered in \secref{sec:T5specialfields} and compared with known expressions from the literature. Finally, summary and conclusions are provided in \secref{sec:conclusion}.

From now on we use natural units $\hbar = c = 1$ and Heaviside-Lorentz units for charge [$\alpha = \nfrac{e^2}{(4\pi)} \approx \nfrac{1}{137}$ denotes the fine-structure constant], the notation agrees with \cite{meuren_polarization_2013}.

\section{Neutrino-photon interactions inside strong laser fields}
\label{sec:neutrinoradiation}

\begin{figure}
\centering
\hfill
\begin{minipage}[c]{0.45\columnwidth}
\centering
\includegraphics{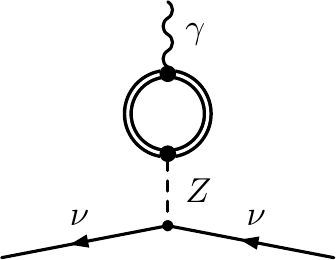}
\end{minipage}
\hfill
\begin{minipage}[c]{0.45\columnwidth}
\centering
\includegraphics{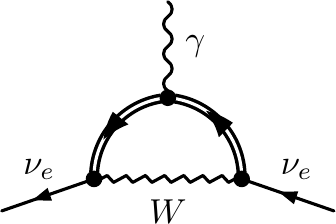}
\end{minipage}
\hfill
\\[1em]
\hfill
\begin{minipage}[c]{0.45\columnwidth}
\small\textbf{a}) $Z$ boson exchange
\end{minipage}
\hfill
\begin{minipage}[c]{0.45\columnwidth}
\small\textbf{b}) $W$ boson exchange
\end{minipage}
\hfill
\\[1.5em]
\begin{minipage}[c]{0.7\columnwidth}
\centering
\includegraphics{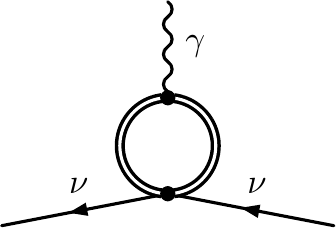}
\\[1.em]
\small\textbf{c}) Local limit (Fermi description)
\end{minipage}
\caption{\label{fig:neutrinophotonvertex}Neutrino-photon interaction vertex. \textbf{a}) The electron-positron loop interacts via the neutral current with all neutrino flavor states. \textbf{b}) Electron neutrinos also couple via the charged current to electrons and positrons. \textbf{c}) In the local limit (exchanged momentum much smaller than the weak gauge-boson mass) the effective four-point Fermi interaction is obtained. The double lines denote dressed electron and positron states which take into account exactly the laser field (time axis from right to left).}
\end{figure}

As neutrinos are neutral particles, their interaction with photons must be mediated by loop diagrams which contain electrically charged particles (see \figref{fig:neutrinophotonvertex}). At the loop level the quantization of the electroweak sector of the standard model involves ``unphysical'' degrees of freedom, i.e. particles which appear only in loops but not as free, asymptotic states \cite{bardin_standard_1999,donoghue_dynamics_1994}. These are the unphysical scalar Higgs particles, present if the calculation is performed in a renormalizable gauge (from the Higgs doublet, which consists of four scalar fields, only one degree of freedom corresponds to the physically observable Higgs particle) and the Feynman-Faddeev-Popov ghosts, which appear in the quantization of a nonabelian gauge theory \cite{pokorski_gauge_2000,cheng_gauge_1982}. Therefore, the complete set of Feynman rules for the electroweak sector of the standard model after symmetry breaking is rather large \cite{aoki_electroweak_1982,bohm_oneloop_1986,hollik_radiative_1990,denner_techniques_1993}. Fortunately, the leading-order contribution (with respect to the electroweak mass scale) to the neutrino-photon coupling inside a plane-wave background field is given by only two diagrams, which are shown in \figref{fig:neutrinophotonvertex} (see also \cite{dvornikov_electromagnetic_2004,gvozdev_radiative_1996}). 

Due to the existence of neutrino oscillations we know that neutrinos have a finite mass \cite{olive_review_2014,bilenky_introduction_2010,avignone_double_2008,giunti_fundamentals_2007,mohapatra_massive_2004,hohler_neutrino_2003,bilenky_massive_1987,kayser_quantum_1981}. The left-handed neutrino mass eigenstates $\nu_{r L}$ ($r=1,2,3$) are related by the unitary Pontecorvo-Maki-Nakagawa-Sakata (PMNS) matrix $\PMNS_{\alpha r}$ (which is also simply called neutrino mixing matrix) $\nu_{\alpha L} = \PMNS_{\alpha r} \, \nu_{r L}$ \cite{pontecorvo_mesonium_1957,pontecorvo_inverse_1958,maki_remarks_1962,olive_review_2014} to the left-handed flavor neutrino eigenstates $\nu_{\alpha L}$ ($\alpha=e,\mu,\tau$).  As neutrinos are produced via the charged current as left-handed flavor eigenstates, the nature of their right-handed component (required for the construction of a mass term in the Lagrangian) is not determined so far, i.e. the neutrino could be either a Dirac or a Majorana particle. At high energies, however, the neutrino mass can usually be neglected and with a reasonable experimental precision it is not possible to distinguish between Dirac and Majorana neutrinos. Correspondingly, we can assume in the following that the neutrino is a massless, left-handed Dirac particle as originally postulated in the standard model.

\subsection{Lagrangian density}

After electroweak-symmetry breaking the Lagrangian density, which describes the interaction between the photon field $\mc{A}^\mu$ and the various fermion fields $\psi_f$, is given by \cite{olive_review_2014,donoghue_dynamics_1994}
\begin{gather}
\label{eqn:sm_elmagcurrent}
\begin{gathered}
\mc{L}_{\mathrm{L}}^{\mathrm{EM}}
=
e \mc{A}_\mu J^\mu_{\mathrm{EM}},
\\
J^\mu_{\mathrm{EM}} = -\bar{\psi}_e \gamma^\mu \psi_e + \frac{2}{3} \bar{\psi}_u \gamma^\mu \psi_u - \frac{1}{3} \bar{\psi}_d \gamma^\mu \psi_d + \ldots
\end{gathered}
\end{gather}
[the index $f=e,\mu,\tau,\ldots$ labels the type of fermion field (quarks and leptons), Dirac spinor indices are suppressed (note that we use the convention $e<0$)]. Correspondingly, we obtain 
\begin{gather}
\label{eqn:sm_Zcurrent}
\begin{gathered}
\mc{L}_{\mathrm{L}}^{\mathrm{Z}}
=
-\frac{g}{2\cos\theta_W}  Z_\mu J^\mu_{\mathrm{Z}},
\\
J^\mu_{\mathrm{Z}} = \bar{\psi}_f \big[g_v^{(f)} \gamma^\mu + g_a^{(f)} \gamma^\mu \gamma^5\big] \psi_f
\end{gathered}
\end{gather}
for the interaction with the $Z$ boson field $Z^\mu$. Here $g=g_2$ and $g'=g_1= g\tan\theta_W $ are the fundamental coupling constants for weak isospin and hypercharge, respectively, which are (after the symmetry breaking) related to the electron charge $e$ and the Fermi constant $G_F$ by $-e = g \sin\theta_W$ and $G_F = \nfrac{(g^2 \sqrt{2})}{(8M_W^2)} \approx \unit[1.17\times 10^{-5}]{GeV^{-2}}$, respectively (at tree level the gauge-boson masses obey $M_W = M_Z\cos\theta_W$, $\theta_W$ is called the weak mixing or Weinberg angle). The constants $g_v^{(f)}$ and $g_a^{(f)}$ depend on the type of fermion. For the charged leptons we obtain $g_v^{(e,\mu,\tau)} = -\nfrac{1}{2} + 2 \sin^2\theta_W$ and $g_a^{(e,\mu,\tau)} = -\nfrac{1}{2}$, for the (massless) neutrinos $g_v^{(\nu_e,\nu_\mu,\nu_\tau)} = g_a^{(\nu_e,\nu_\mu,\nu_\tau)} =  \nfrac{1}{2}$ and for the quarks $g_v^{(u,c,t)} = \nfrac{1}{2}-(\nfrac{4}{3})\sin^2\theta_W$ $g_v^{(d,s,b)} = -\nfrac{1}{2}+(\nfrac{2}{3})\sin^2\theta_W$, $g_a^{(u,c,t)} = - g_a^{(d,s,b)} = \nfrac{1}{2}$ [note that we use the same notation for $\gamma^5$ as in \cite{landau_quantum_1981} and \cite{donoghue_dynamics_1994}, i.e. the projection operators $P_L$ for the left-handed and $P_R$ for the right-handed component are given by $P_L = \nfrac{(\one+\gamma^5)}{2}$ and $P_R = \nfrac{(\one-\gamma^5)}{2}$, respectively].

Finally, the Lagrangian density, which describes the interaction between the complex $W$ boson field $W^+_\mu$ [the plus is part of the symbol name, we also define $W^-_\mu = (W^+_\mu)^\dagger$] and the first lepton generation can be written as
\begin{gather}
\label{eqn:sm_Wcurrent}
\begin{gathered}
\mc{L}_{e}^{\mathrm{W}}
=
-\frac{g}{2\sqrt{2}} [W^+_\mu J_{\mathrm{W},e}^\mu + W^-_\mu (J_{\mathrm{W},e}^\mu)^\dagger],
\\
J_{\mathrm{W},e}^\mu = \bar{\psi}_{\nu_e} \gamma^\mu (\one + \gamma^5) \psi_e.
\end{gathered}
\end{gather}

From Eqs.\,(\ref{eqn:sm_elmagcurrent})-(\ref{eqn:sm_Wcurrent}) one obtains the interaction vertices between the fermions and the electroweak gauge fields of the standard model \cite{donoghue_dynamics_1994,aoki_electroweak_1982}; they contain both vector ($\gamma^\mu$) and axial-vector ($\gamma^\mu \gamma^5$) couplings. 

After quantization the propagators for the weak gauge bosons are (in position space and Feynman gauge) given by \cite{donoghue_dynamics_1994}
\begin{multline}
\label{eqn:WZbosonpropagator}
\I G_{Z,W}^{\mu\nu}(x-y) 
\\= 
\int \frac{d^4p}{(2\pi)^4} \, \frac{-\I g^{\mu\nu}}{p^2-M_{Z,W}^2+\I0} \, e^{-\I p(x-y)}.
\end{multline}
If the exchanged momenta are much smaller than the weak mass scale $M_{Z,W} \sim \unit[100]{GeV}$, one can neglect the momentum dependence in the denominator in Eq. (\ref{eqn:WZbosonpropagator}) (local limit). After taking the momentum integrals, the propagators are then given by
\begin{gather}
\I G_{Z,W}^{\mu\nu}(x-y)
=
\I \frac{g^{\mu\nu}}{M_{Z,W}^2} \delta^4(x-y)
\end{gather}
(see \figref{fig:neutrinophotonvertex}c). Physically, this means that the $Z$ and the $W$ boson are too heavy to propagate a significant distance and we obtain essentially Fermi's description for the weak force \cite{fermi_tentativo_1934,fermi_versuch_1934}.

\subsection{Plane-wave background fields}
\label{sec:planewavefields}

In the following we will consider an external plane-wave laser field described by the field tensor
\begin{gather}
F^{\mu\nu}(\phi)
=
\del^\mu A^\nu(\phi) - \del^\nu A^\mu(\phi) 
=
\sum_{i=1,2} f_i^{\mu\nu} \psi'_i(\phi),
\end{gather}
$\phi = kx$ (the prime denotes the derivative with respect to the argument). Here, $A^\mu(\phi)$ denotes the four-potential and
\begin{gather}
f_i^{\mu\nu} = k^\mu a_i^\nu - k^\nu a_i^\mu,\\
f^{\mu}_{i\,\rho} f_{j}^{\rho\nu} = -\delta_{ij} a_i^2\, k^\mu k^\nu,
\quad
k_\mu f_i^{\mu\nu} = 0
\end{gather}
($k^2 = ka_i = a_1 a_2 = 0$). We also introduce the integrated field tensor
\begin{gather}
\Ftilde^{\mu\nu}(\phi)
= 
\int^{\phi}_{-\infty} d\phi'\, F^{\mu\nu}(\phi')
=
\sum_{i=1,2} f_i^{\mu\nu} \psi_i(\phi).
\end{gather}
Correspondingly, the scalar functions $\psi_i(\phi)$ describe the shape of the laser field. They are arbitrary (differentiable) functions, restricted only by the physical requirement that the external field is of finite extent and has no dc component [i.e., $\psi_i(\pm\infty)=\psi'_i(\pm \infty)=0$, with $\psi_i(\phi)$, $\psi'_i(\phi)$ vanishing fast enough at infinity]. Furthermore, we assume (without restriction) that $\abs{\psi_i(\phi)},\abs{\psi'_i(\phi)} \lesssim 1$. This implies that the field strength is measured by the following (classical) intensity parameters
\begin{gather}
\xi_i = \frac{\abs{e}}{m} \sqrt{-a_i^2}.
\end{gather}
Calculations with plane-wave background fields become particularly transparent if light-cone coordinates are used \cite{dirac_forms_1949,neville_quantum_1971,mitter_quantum_1975}
\begin{gather}
\label{eqn:paircreation_lcc}
v^\lminus = vk,
\quad
v^\lplus = v\bar{k},
\quad
v^\lone	=  ve_1,
\quad
v^\ltwo	=  ve_2
\end{gather}
($v^\mu$ is an arbitrary four-vector, $\lone$ and $\ltwo$ are also summarized as $\lperp$). Here, we require that the four four-vectors $k^\mu$, $\bar{k}^\mu$, $e_1^\mu$ and $e_2^\mu$ form a light-cone basis  [see App.\,C and Eq.\,(32) of Ref.\,\cite{meuren_polarization_2013} and Ref.\,\cite{meuren_semiclassical_2015}]. 

More details can be found in Refs. \cite{dittrich_probingquantum_2000,fradkin_quantum_1991,ritus_1985,mitter_quantum_1975} and in the recent review articles \cite{battesti_magnetic_2013,di_piazza_extremely_2012,ehlotzky_fundamental_2009,mourou_optics_2006,marklund_nonlinear_2006}.

\subsection{Z boson exchange}

In the local limit the matrix element for the emission of a photon with four-momentum $q^\mu$ and polarization four-vector $\eps^\mu$ by a neutrino due to $Z$ boson exchange (see \figref{fig:neutrinophotonvertex}a) is given by \cite{ioannisian_cherenkov_1997}
\begin{multline}
\label{eqn:neutrinophotonemission_matrixelementZboson}
\I\mathfrak{M}_Z(p',q;p) = \frac{2 G_F}{\sqrt{2}} \, \bar{u}_{\nu,p'} \gamma_\mu P_L u_{\nu,p} 
\\
\times \frac{1}{e}\Big[g_v^{(e)} T^{\mu\nu}(p-p',q) +  g_a^{(e)} T_5^{\mu\nu}(p-p',q) \Big] \eps_\nu^*
\end{multline}
(we use the same conventions for matrix elements as in \cite{meuren_polarization-operator_2015}). Here $u_{\nu,p}$ and $u_{\nu,p'}$ are the Dirac spinors for the incoming neutrino with four-momentum $p^\mu$ and the outgoing neutrino with four-momentum $p'^\mu$, respectively. Furthermore, $T^{\mu\nu}(q_1,q_2)$ denotes the polarization tensor (see \figref{fig:T5vsT} and \cite{gies_laser_2014,dinu_photon_2014,dinu_vacuum_2014,meuren_polarization_2013,baier_interaction_1975,becker_vacuum_1975})
\begin{multline}
\label{eqn:polarizationoperator}
\T^{\mu\nu}(q_1,q_2)
=
\int  \frac{d^4p\, d^4p'}{(2\pi)^8} 
\tr 
\, 
\Gamma^\mu(p',q_1,p)
\\ \times  \frac{\s{p} + m}{p^2-m^2+i0} \, 
\Gamma^\nu(p,-q_2,p')
\frac{\s{p}' + m}{p'^2-m^2+i0}
\end{multline}
and $\T_5^{\mu\nu}(q_1,q_2)$ the axial-vector--vector current-coupling tensor
\begin{multline}
\label{eqn:axialvectorcouplingtensor}
\T_5^{\mu\nu}(q_1,q_2)
=
\int  \frac{d^4p\, d^4p'}{(2\pi)^8} 
\tr 
\, 
\Gamma^\mu(p',q_1,p) \gamma^5
\\ \times  \frac{\s{p} + m}{p^2-m^2+i0} \, 
\Gamma^\nu(p,-q_2,p')
\frac{\s{p}' + m}{p'^2-m^2+i0},
\end{multline}
which will be calculated in \secref{sec:T5} [the final result is given in \eqref{eqn:axialvectorcouplingtensorfinal}; for the definition of the dressed vertex $\Gamma^\mu$ see \appref{sec:dressedvertexappendix}]. Note that $\T_5^{\mu\nu}(q_1,q_2)$ is actually a pseudo-tensor and that in our definition the electron charge $e$ is taken as the coupling constant for both vertices [this is the reason for the prefactor $\nfrac{1}{e}$ appearing in \eqref{eqn:neutrinophotonemission_matrixelementZboson}]. Furthermore, the appearance of $g_v^{(e)}$ and $g_a^{(e)}$ in \eqref{eqn:neutrinophotonemission_matrixelementZboson} is related to the electron-positron loop and is independent of the neutrino species which interacts with the photon.

Despite the fact that the electron and the positron are the lightest charged fermions, also the muon, the tau and the various quarks contribute to the loop which couples the photon with the $Z$ boson (see \figref{fig:neutrinophotonvertex}a). To calculate the contribution of the other charged fermions to this loop, the electron (positron) mass and charge appearing in $\frac{1}{e}\T^{\mu\nu}(q_1,q_2)$ and $\frac{1}{e}\T_5^{\mu\nu}(q_1,q_2)$ must be replaced accordingly and the $Z$ boson coupling constants $g_{v,a}^{(e)}\to g_{v,a}^{(f)}$ adjusted. As discussed in the introduction, the nonlinear interaction with the background field can be neglected for fermions with a mass well above the electron (positron) scale (for reasonable field strengths of the background field). However, the contribution of all fermions in each generation is needed for the cancellation of the axial-vector anomaly. As the anomalous contribution to $\frac{1}{e}T_5^{\mu\nu}(q_1,q_2)$ is independent of the fermion mass and proportional to the square of the electric charge [at one loop in the presence of a plane-wave, see \eqref{eqn:axialvectorcouplingtensorfinal_anomaly}], this cancellation (for each fermion generation) can be seen from the relation
\begin{gather}
\frac{1}{2}\big[-(-1)^2 + 3 \lb\nfrac{2}{3}\rb^2 - 3 \lb-\nfrac{1}{3}\rb^2\big]  = 0
\end{gather}
(all gauge-symmetry anomalies must cancel in the standard model, otherwise it would be nonrenormalizable \cite{gross_effect_1972}).

\subsection{W boson exchange}

For electron neutrinos also the exchange of a $W$ boson contributes to the photon-emission matrix element (see \figref{fig:neutrinophotonvertex}b). Applying the local limit for the $W$ boson propagator, we obtain the following expression for the matrix element
\begin{multline}
\label{eqn:neutrinophotonemission_matrixelementWbosonA}
\I\mathfrak{M}_W(p',q;p) = \frac{4 G_F}{\sqrt{2}} \, \int \frac{dp_1^4 dp_2^4}{(2\pi)^8} \int d^4x \, e^{-\I (p-p')x}
\\\times
\bar{u}_{\nu,p'} \gamma^\rho P_L  \, M^\nu(p_1,p_2,q;x)\eps_\nu^* \, \gamma_\rho P_L u_{\nu,p},
\end{multline}
where 
\begin{multline}
M^\nu(p_1,p_2,q;x) = \I E_{p_1,x} \frac{\s{p}_1+m}{p_1^2-m^2+\I0} 
\\\times \Gamma^\nu(p_1,-q,p_2) \frac{\s{p}_2+m}{p_2^2-m^2+\I0} \bar{E}_{p_2,x}.
\end{multline}
Using the identities given in \appref{sec:gammaalgebraappendix} for an arbitrary $4\times 4$ matrix $\Gamma$ in spinor space, we obtain
\begin{gather}
\gamma^\rho P_L \Gamma \gamma_\rho P_L
=
-
2r_\mu \gamma^\mu P_L,
\end{gather}
where
\begin{gather}
r_\mu = \tfrac12 \tr P_R \gamma_\mu  \Gamma = \tfrac12 \tr \gamma_\mu  P_L \Gamma.
\end{gather}
Therefore, we can write the matrix element for the $W$ boson exchange diagram as [see \eqref{eqn:neutrinophotonemission_matrixelementWbosonA}] \cite{ioannisian_cherenkov_1997}
\begin{multline}
\label{eqn:neutrinophotonemission_matrixelementWbosonB}
\I\mathfrak{M}_W(p',q;p) = \frac{2 G_F}{\sqrt{2}} \, \bar{u}_{\nu,p'} \gamma_\mu P_L u_{\nu,p} 
\\
\times \frac{1}{e}\Big[T^{\mu\nu}(p-p',q) +  T_5^{\mu\nu}(p-p',q) \Big] \eps_\nu^*
\end{multline}
(note that it only contributes for electron neutrinos). In the local limit it has the same structure as the one for the $Z$ boson exchange given in \eqref{eqn:neutrinophotonemission_matrixelementZboson}. The anomaly, however, must drop also for this diagram if one first performs the calculations by employing the full $W$ boson propagator in Eq. (\ref{eqn:WZbosonpropagator}) \cite{ioannisian_cherenkov_1997}.

\section{Current-coupling tensor}
\label{sec:T5}

In the previous section it was shown how the axial-vector--vector current-coupling tensor $\T_5^{\mu\nu}(q_1,q_2)$ (see \figref{fig:T5vsT}) arises in the calculation of neutrino-photon interactions inside strong laser fields. Now, we will examine $\T_5^{\mu\nu}(q_1,q_2)$ closer and derive a compact triple-integral representation for it. After applying the Feynman rules \cite{di_piazza_extremely_2012,dittrich_probingquantum_2000,fradkin_quantum_1991,ritus_1985,mitter_quantum_1975}, we obtain the following expression [see \eqref{eqn:axialvectorcouplingtensor}]
\begin{multline}
\T_5^{\mu\nu}(q_1,q_2)
=
\int  \frac{d^4p\, d^4p'}{(2\pi)^8} 
\tr 
\, 
\Gamma^\mu(p',q_1,p) \gamma^5
\\ \times  \frac{\s{p} + m}{p^2-m^2+i0} \, 
\Gamma^\nu(p,-q_2,p')
\frac{\s{p}' + m}{p'^2-m^2+i0}
\end{multline}
(for the definition of the dressed vertex $\Gamma^\rho$ see \appref{sec:dressedvertexappendix}). Note that $\T_5^{\mu\nu}(q_1,q_2)$ is equivalent to the tensor
\begin{multline}
\widetilde{\T}_5^{\mu\nu}(q_1,q_2)
=
\int  \frac{d^4p\, d^4p'}{(2\pi)^8} 
\tr 
\, 
\Gamma^\mu(p',q_1,p)
\\ \times  \frac{\s{p} + m}{p^2-m^2+i0} \, 
\Gamma^\nu(p,-q_2,p') \gamma^5\,
\frac{\s{p}' + m}{p'^2-m^2+i0},
\end{multline}
which is obtained by interchanging the vector- and the axial-vector current vertex [$\widetilde{\T}_5^{\mu\nu}(q_1,q_2) = \T_5^{\nu\mu}(-q_2,-q_1)$].

The current-coupling tensor $\T_5^{\mu\nu}(q_1,q_2)$, which we consider here, differs from the polarization tensor $\T^{\mu\nu}(q_1,q_2)$ [see \eqref{eqn:polarizationoperator}] only by the insertion of $\gamma^5$ (i.e. by the trace). Hence, the calculation of $\T_5^{\mu\nu}(q_1,q_2)$ is related to the one of $\T^{\mu\nu}(q_1,q_2)$ carried out in \cite{meuren_polarization_2013} and we will focus here mainly on the differences between both derivations. At first sight one may think that the small modification of the trace should only affect the technical details of the calculation. However, it is responsible for several important qualitative changes like the appearance of the ABJ anomaly in $\T_5^{\mu\nu}(q_1,q_2)$, which we will discuss now in detail.

An important consequence of the additional $\gamma^5$ in the trace of $\T_5^{\mu\nu}(q_1,q_2)$ is the fact that only an odd number of couplings to the background field are allowed if the background field is treated perturbatively (this follows from a generalization of Furry's theorem, see e.g. \cite{nishijima_generalized_1951}; for the polarization tensor only an even number of couplings is possible). Accordingly, the tensor structure of $\T_5^{\mu\nu}(q_1,q_2)$ is different from thatof $\T^{\mu\nu}(q_1,q_2)$ [see \eqref{eqn:axialvectorcouplingtensordecomposition}] and the vacuum contribution to $\T_5^{\mu\nu}(q_1,q_2)$ vanishes (see \figref{fig:anomaly}). Furthermore, we will see that no infinities are encountered and $\T_5^{\mu\nu}(q_1,q_2)$ does not need to be regularized.
 
In order to determine $\T_5^{\mu\nu}(q_1,q_2)$ we insert the dressed vertex (see \appref{sec:dressedvertexappendix}) into \eqref{eqn:axialvectorcouplingtensor} [we will denote the vertex integrals associated with~$\Gamma^\mu(p',q_1,p)$ and~$\Gamma^\nu(p,-q_2,p')$ by~$d^4x$ and~$d^4y$, respectively] and obtain 
\begin{multline}
\label{eqn:axialvectorcouplingtensorB}
\T_5^{\mu\nu}(q_1,q_2)
=
4 \, (-ie)^2 \int  \frac{d^4p\, d^4p'}{(2\pi)^8} \int d^4x d^4y\,
 \\ \times \,
\frac{\frac14 \tr \big[ \cdots \big]_5^{\mu\nu}}{(p^2-m^2+i0)(p'^2-m^2+i0)} e^{iS_\T},
\end{multline}
where
\begin{multline}
\label{eqn:sfqed_polarizationoperatorphasestructure}
\I S_\T
= 
\I(p'-p-q_1)x 
+
\I(p-p'+q_2)y
\\+ 
\I\int_{ky}^{kx} d\phi'\, \bigg[ \frac{e p_\mu p'_\nu \Ftilde^{\mu\nu}(\phi')}{(kp)(kp')} 
\\
+ \frac{e^2(kp-kp')}{2(kp)^2(kp')^2} p_\mu p'_\nu \Ftilde^{2\mu\nu}(\phi') \bigg]
\end{multline}
and the trace in \eqref{eqn:axialvectorcouplingtensorB} is given by
\begin{widetext}
\begin{multline}
\label{eqn:sfqed_axialvectorcouplingtensortraceA}
\frac14 \tr \big[ \cdots \big]_5^{\mu\nu}
=
\frac14 \tr 
\big[ \gamma_\alpha a^{\alpha\mu} + i\gamma_\alpha \gamma^5 b^{\alpha\mu} \big] \gamma^5
(\s{p} + m) \, 
\big[ \gamma_\beta c^{\beta\nu} + i\gamma_\beta \gamma^5 d^{\beta\nu} \big]
(\s{p}' + m)
=
  \I m^2[(b^{\alpha\mu} c_\alpha^{\phantom{\alpha}\nu}) -(a^{\alpha\mu}d_\alpha^{\phantom{\alpha}\nu})]
\\ 
- \I (pp')[(a^{\alpha\mu}d_\alpha^{\phantom{\alpha}\nu})  + (b^{\alpha\mu} c_\alpha^{\phantom{\alpha}\nu})]
+ \I (p_\alpha p'_\beta + p_\beta p'_\alpha) (b^{\alpha\mu} c^{\beta\nu} + a^{\alpha\mu} d^{\beta\nu})
- \I \eps_{\rho\sigma\alpha\beta} p^\rho p'^\sigma (b^{\alpha\mu} d^{\beta\nu} - a^{\alpha\mu} c^{\beta\nu}),
\end{multline}
\end{widetext}
where
\begin{gather}
\begin{aligned}
a^{\alpha\mu} 
&=
G^{\alpha\mu}(kp',kp;kx),&
c^{\beta\nu}
&=
G^{\beta\nu}(kp,kp';ky),\\
b^{\alpha\mu}
&= 
G_5^{\alpha\mu}(kp',kp;kx),& 
d^{\beta\nu} 
&=
G_5^{\beta\nu}(kp,kp';ky)
\end{aligned}
\end{gather}
[the additional $\gamma^5$ in \eqref{eqn:sfqed_axialvectorcouplingtensortraceA} exchanges the vector and the axial-vector current in comparison with the polarization tensor].

Next, we employ the proper time parametrization for the propagators
\begin{multline}
\label{eqn:sfqed_polarizationoperatorschwingerpropagator}
\frac{1}{p^2-m^2+\I0} \frac{1}{p'^2-m^2+\I0}
=
(-\I)^2 \int_0^\infty ds \, dt \, 
\\ \times \, \exp \lsb \I(p^2-m^2+i0)s + \I(p'^2-m^2+i0)t \rsb
\end{multline}
(the pole prescription $i0$ will be dropped in the exponents and can be restored by substituting $m^2\to m^2-i0$). In the final representation we change the proper-time integrals in the following way
\begin{gather}
\label{eqn:stintegraltransform}
\int_0^\infty ds \, dt  \, f(s,t)
=
\frac12 \int_{-1}^{+1} dv \int_0^\infty d\tau \, \tau \tilde{f}(\tau,v)
\end{gather}
(note that terms odd in $v$ in the resulting function $\tilde{f}(\tau,v)$ can be dropped), where 
\begin{gather}
\label{eqn:tauvmudefinition}
\tau = s+t,
\quad
v = \frac{s-t}{s+t},
\quad
\mu = \frac{st}{s+t}
= 
\frac14 \tau (1-v^2).
\end{gather}
After including the parametrization (\ref{eqn:sfqed_polarizationoperatorschwingerpropagator}) in Eq. (\ref{eqn:axialvectorcouplingtensorB}) and by adding the source terms $\I p_\mu j^\mu + \I p'_\mu j'^\mu$ to the resulting phase in the same equation, we apply the replacements
\begin{gather}
\label{eqn:momentumreplacementtrace}
p^\mu \longrightarrow (-\I)\del_j^\mu,
\quad
p'^\mu \longrightarrow (-\I)\del_{j'}^\mu
\end{gather}
to the trace given in \eqref{eqn:sfqed_axialvectorcouplingtensortraceA}, use modified light-cone coordinates [see Eq. (C11) in \cite{meuren_polarization_2013}] and define
\begin{gather}
\label{eqn:sfqed_polarizationoperatorintegraldefinitions}
\begin{aligned}
\uplambda^\mu 
&=
- \frac{m (kq)}{(kp)(kp')} \sum_{i=1,2} \xi_i \Lambda_i^\mu \int_{ky}^{kx} d\phi' \, \psi_i(\phi'),\\
\Uplambda
&=  
-\frac{m^2 (kq)}{2(kp)(kp')} \sum_{i=1,2} \xi^2_i \int_{ky}^{kx} d\phi' \,  \psi_i^2(\phi'),
\end{aligned}
\end{gather}
\begin{gather}
\label{eqn:defIJintegrals}
\begin{aligned}
I_i 
&=
-\frac{1}{2kq \mu} \int_{ky}^{kx} d\phi' \, \psi_i(\phi'),\\
J_i 
&=
-\frac{1}{2kq \mu} \int_{ky}^{kx} d\phi' \, \psi^2_i(\phi')
\end{aligned}
\end{gather}
(see \appref{sec:basisvectorsappendix} for more details). Finally, we obtain the following structure for the current-coupling tensor
\begin{multline}
\label{eqn:T5withpropertimeintegrals}
T_5^{\mu\nu}(q_1,q_2) 
= 
4 (-\I e)^2  \int  \frac{d^4p\, d^4p'}{(2\pi)^8} \int d^4x d^4y 
\\
\times (-\I)^2 \int_0^\infty ds dt \, \frac{1}{4} \tr [\cdots]^{\mu\nu}_5 e^{\I S'_T},
\end{multline}
where the full phase including the proper-time exponents and the source terms is given by
\begin{gather}
\label{eqn:polopinitialphasedefinition}
\begin{aligned}
S'_T 		&=  \tilde{S}_T + p_\mu j^\mu + p'_\mu j'^\mu,
\\
\tilde{S}_T &= (p^2-m^2) s + (p'^2-m^2) t + S_T
\end{aligned}
\end{gather}
(if no explicit argument is present, the prime is a part of the symbol name and does not indicate a derivative). After taking most of the integrals we obtain the result [see Eq. (64) in \cite{meuren_polarization_2013}]
\begin{multline}
\label{eqn:T5afterintegralstaken}
\T_5^{\mu\nu}(q_1,q_2)
= 
-2i\pi e^2\, \delta^{(\lminus,\lperp)}(q_1-q_2) 
\int_0^\infty ds \, dt 
\\ \times\,
\int_{-\infty}^{+\infty} dx^\lminus \,
\frac{1}{(s+t)^2} \frac14 \tr \lsb \cdots \rsb^{\mu\nu}_5  e^{iS'_\T} \Big|_{j=j'=0},
\end{multline}
where $x^\lminus = kx = \phi$ and the trace is given in \eqref{eqn:sfqed_axialvectorcouplingtensortraceA} with the replacement in \eqref{eqn:momentumreplacementtrace} and 
\begin{multline}
\label{eqn:polarizationoperatorfinalphase}
iS'_\T
=
i \big[ (q_2^\lplus -q_1^\lplus) x^\lminus 
- 
m^2 (s+t)
+ 
\frac{st}{s+t} q_2^2 
\\-
\frac{1}{s+t} (t \, q_2j - s\, q_2j')
- 
\frac{1}{4(s+t)} (j+j')^2 
\\-
\frac{1}{2(s+t)} (j+j')\uplambda
-
\frac{1}{4(s+t)} \uplambda^2
+  
\Uplambda \big]
\end{multline}
[note that, since no confusion can arise, we use the same symbol for the phase before and after the mentioned integrals are taken, see Eqs. (\ref{eqn:T5withpropertimeintegrals}) and (\ref{eqn:polopinitialphasedefinition}) and Eqs. (\ref{eqn:T5afterintegralstaken}) and (\ref{eqn:polarizationoperatorfinalphase})].

Due to the momentum-conserving delta functions in \eqref{eqn:T5afterintegralstaken} we will simply write $q^\mu$ whenever $q_1^\mu$ and $q_2^\mu$ can be used interchangeably. 

To obtain a symmetric expression with respect to the external momenta $q^\mu_1$ and $q^\mu_2$ we will perform below the shift 
\begin{gather}
\label{eqn:symmetricintegralshift}
z^\lminus = x^\lminus + \mu q^\lminus
\end{gather}
and use $z^\lminus$ as integration variable.

\subsection{Ward-Takahashi identity}
\label{sec:T5wardidentity}

According to Noether's theorem the gauge invariance of the QED Lagrangian implies electric charge conservation. More specifically, if the spinor field $\psi(x)$ obeys the Dirac equation
\begin{gather}
[\I\s{\del} - e\s{A}(x) - m]\psi(x) = 0
\end{gather} 
[here $A^\mu(x)$ denotes the classical background field], the vector current is conserved
\begin{gather}
\del_\mu j^\mu(x) = 0,
\quad
j^\mu(x) = \bar{\psi}(x) \gamma^\mu \psi(x).
\end{gather}
After changing to momentum space
\begin{gather}
j^\mu_q 
=
\int d^4x \, e^{-\I qx} j^\mu(x),
\end{gather}
the current conservation law is expressed by
\begin{gather}
\int d^4x \, e^{-\I qx} [\del_\mu j^\mu(x)]
=
\I q_\mu j^\mu_q=0.
\end{gather}
Therefore, one expects that the polarization tensor $T^{\mu\nu}(q_1,q_2)$ obeys the following homogeneous Ward-Takahashi identity 
\begin{gather}
\label{eqn:polopwardidentity}
q_{1\mu} T^{\mu\nu}(q_1,q_2)
=
0,
\quad
T^{\mu\nu}(q_1,q_2) q_{2\nu}
=
0,
\end{gather}
which is indeed the case \cite{meuren_polarization_2013}. 

Correspondingly, the divergence of the axial-vector current 
\begin{gather}
j_5^\mu(x) = \bar{\psi}(x) \gamma^\mu\gamma^5  \psi(x)
\end{gather}
should be given by
\begin{gather}
\label{eqn:T5currentdivergenceclassical}
\del_\mu j_5^\mu(x) = 2m\I \, \bar{\psi}(x)  \gamma^5  \psi(x).
\end{gather}

After applying \eqref{eqn:algebraicidentitydressedvertex} to the definition of $\T_5^{\mu\nu}(q_1,q_2)$ [see \eqref{eqn:axialvectorcouplingtensor}] we obtain the following Ward-Takahashi identity for the tensor $T_5^{\mu\nu}(q_1,q_2)$
\begin{gather}
\label{eqn:T5wardidentity}
\begin{aligned}
q_{1\mu} \T_5^{\mu\nu}(q_1,q_2) &= \phantom{-} T_5^\nu(q_1,q_2) + \mathfrak{T}_5^\nu(q_1,q_2),
\\
\T_5^{\mu\nu}(q_1,q_2) q_{2\nu} &= -\mathfrak{T}_5^\mu(-q_2,-q_1),
\end{aligned}
\end{gather}
where we defined
\begin{multline}
\label{eqn:qonecontractionwithtensor_massterm}
\T_5^\nu(q_1,q_2) 
= 
2m \int  \frac{d^4p\, d^4p'}{(2\pi)^8} 
\tr 
\, 
I(p',q_1,p) \gamma^5
\\ \times  \frac{\s{p} + m}{p^2-m^2+i0} \, 
\Gamma^\nu(p,-q_2,p')
\frac{\s{p}' + m}{p'^2-m^2+i0}
\end{multline}
and the so-called anomalous contribution
\begin{multline}
\label{eqn:qonecontractionwithtensor_anomalous}
\mathfrak{T}_5^\nu(q_1,q_2) 
= 
\int  \frac{d^4p\, d^4p'}{(2\pi)^8} 
\\
\begin{aligned}
\times \Big[&\tr \,
\Gamma^\nu(p,-q_2,p') I(p',q_1,p) \gamma^5 \frac{\s{p} + m}{p^2-m^2+i0}  
\\-
&\tr \, I(p',q_1,p) \Gamma^\nu(p,-q_2,p') \gamma^5 \frac{\s{p}' + m}{p'^2-m^2+i0} \, \Big].
\end{aligned}
\end{multline}
Furthermore, we introduced the dressed scalar vertex $I(p',q,p)$ [see \eqref{eqn:sfqed_Ivertexdefinition}].

Based on \eqref{eqn:T5currentdivergenceclassical} one expects that the anomalous contribution to the Ward-Takahashi identity vanishes. If the relations given in \eqref{eqn:IGammacontraction} are formally applied to \eqref{eqn:qonecontractionwithtensor_anomalous} as in Sec. II.E of \cite{meuren_polarization_2013}, it looks like this is really the case. However, a closer analysis reveals that the intermediate expressions are divergent and the required formal manipulations cannot be justified. In fact, as shown in \secref{sec:anomaly}, the anomalous contribution does not vanish and is given by
\begin{multline}
\label{eqn:T5anomalyresult_final}
\mathfrak{T}_5^\nu(q_1,q_2) 
= -\I\pi e^2 \delta^{(\lminus,\lperp)}(q_1-q_2) 
\\ \times 4e \int_{-\infty}^{+\infty} dx^\lminus \, e^{\I(q_2^\lplus-q_1^\lplus) x^\lminus} \, q_\mu F^{*\mu\nu}(kx).
\end{multline}
The phenomenon that quantum fluctuations can spoil the results expected from the classical symmetries of the Lagrangian has first been observed in \cite{adler_axial-vector_1969,bell_pcac_1969} and, as we have mentioned, is known as ABJ anomaly (see also \cite{jackiw_anomalies_1969,adler_absence_1969,bardeen_anomalous_1969}, \cite{fujikawa_path-integral_1979} for a discussion using the Feynman path integral and e.g. \cite{peskin_introduction_2008,dittrich_probingquantum_2000,weinberg_quantum_1996} for a textbook discussion).

Unlike the anomalous contribution the calculation of $\T_5^\nu(q_1,q_2)$ [see \eqref{eqn:qonecontractionwithtensor_massterm}] is much less involved. Due to the identity [see \eqref{eqn:sfqed_Ivertexdefinition}]
\begin{multline}
\label{eqn:Igammafivedecomposition}
I(p',q,p) \gamma^5
\\=
-ie \int d^4x \lb \gamma^5 - \frac{1}{2} G_3 \Ftilde^{*\rho\sigma}_x i \sigma_{\rho\sigma} \rb 
e^{iS_\Gamma(p',q,p;x)}
\end{multline}
we only have to change the trace in \eqref{eqn:sfqed_axialvectorcouplingtensortraceA} to
\begin{multline}
\label{eqn:qonecontractiontrace}
\frac14 \tr \big[\cdots\big]_5^\nu
=
\frac14 \tr 
2m\lb \gamma^5 - \frac12 G_3 \Ftilde^{*\rho\sigma}_x i\sigma_{\rho\sigma} \rb
\\\times\,(\s{p} + m) 
\big( \gamma_\beta c^{\beta\nu} + i \gamma_\beta \gamma^5 d^{\beta\nu} \big)
(\s{p}' + m)
\\=
2im^2 \big[ (p-p')_\beta d^{\beta\nu} 
+ 
G_3 (p-p')^\alpha \Ftilde^{*}_{x\alpha\beta} c^{\beta\nu} 
\\- 
G_3 (p+p')^\alpha \Ftilde_{x\alpha\beta} d^{\beta\nu} \big].
\end{multline}
Since the action of the derivatives on $kj$  and $kj'$ gives no contribution, we obtain the replacement rules [see \eqref{eqn:momentumreplacementtrace}]
\begin{gather}
\label{eqn:T5_ppprimereplacementruleswardidentity}
\begin{aligned}
p^\mu 
&\longrightarrow
(-i) \del_j^\mu
\longrightarrow
- \frac{1}{s+t} \lb tq_2^\mu + \frac12 \uplambda^\mu \rb,\\
p'^\mu 
&\longrightarrow
(-i) \del_{j'}^\mu
\longrightarrow
\phantom{-} \frac{1}{s+t} \lb sq_2^\mu - \frac12 \uplambda^\mu \rb,
\end{aligned}
\end{gather}
implying
\begin{gather}
(p-p')^\mu 
\longrightarrow
- q_2^\mu,
\quad
(p+p')^\mu 
\longrightarrow
v q_2^\mu - \frac{1}{\tau} \uplambda^\mu.
\end{gather}
After applying them and noting that terms linear in $v$ vanish [see \eqref{eqn:stintegraltransform}], we can replace the trace in \eqref{eqn:qonecontractiontrace} by
\begin{gather}
\frac14 \tr \big[\cdots\big]_5^\nu
\longrightarrow
2im^2 G_3 \big[ (\Ftilde^*_x q)^\nu - (\Ftilde^*_y q)^\nu \big]
\end{gather}
for the calculation of $\T_5^\nu(q_1,q_2)$ [see \eqref{eqn:qonecontractionwithtensor_massterm}].

\subsection{Tensor structure}

Due to the inhomogeneous Ward-Takahashi identity [see \eqref{eqn:T5wardidentity}] the tensor structure of $\T_5^{\mu\nu}(q_1,q_2)$ is more complicated than that of the polarization tensor $\T^{\mu\nu}(q_1,q_2)$. Using the complete sets $q_1^\mu$, $\Q_1^\mu$, $\Lambda^{*\mu}_1$, $\Lambda^{*\mu}_2$ and $q_2^\nu$, $\Q_2^\nu$, $\Lambda^{*\nu}_1$, $\Lambda^{*\nu}_2$ (see \appref{sec:basisvectorsappendix} for more details) we obtain the following expansion
\begin{multline}
\label{eqn:axialvectorcouplingtensordecomposition}
\T^{\mu\nu}_5
=
\mathfrak{T}^{\mu\nu}_5
+
d_1 \Lambda_1^{*\mu} \Lambda_2^{*\nu}
+
d_2 \Lambda_2^{*\mu} \Lambda_1^{*\nu}
+
d_3 \Lambda_1^{*\mu} \Lambda_1^{*\nu}
\\+
d_4 \Lambda_2^{*\mu} \Lambda_2^{*\nu}
+
d_5 \Q_1^\mu \Q_2^\nu
+
d_6 \Q_1^\mu \Lambda_1^{*\nu}
\\+
d_7 \Q_1^\mu \Lambda_2^{*\nu}
+
d_8 \Lambda_1^{*\mu} \Q_2^\nu
+
d_9 \Lambda_2^{*\mu} \Q_2^\nu
\\+
d_{10} q_1^\mu \Lambda_1^{*\nu}
+
d_{11} q_1^\mu \Lambda_2^{*\nu}
+
d_{12} q_1^\mu \Q_2^\nu,
\end{multline}
where $\mathfrak{T}^{\mu\nu}_5(q_1,q_2)$ contains the contribution from the anomaly, i.e. [see \eqref{eqn:T5anomalyresult_final}]
\begin{multline}
\label{eqn:axialvectorcouplingtensorfinal_anomaly}
\mathfrak{T}^{\mu\nu}_5(q_1,q_2)
= -\I\pi e^2 \delta^{(\lminus,\lperp)}(q_1-q_2) 
\, 4e \, \int_{-\infty}^{+\infty} dx^\lminus \,
\\ \times  e^{\I(q_2^\lplus-q_1^\lplus) x^\lminus}  \, \frac{1}{kq} \lsb k^\mu (qF^{*})^\nu (kx) + (qF^{*})^\mu (kx) k^\nu \rsb.
\end{multline}
As expected from Furry's theorem \cite{nishijima_generalized_1951}, the coefficients $d_1-d_5$ and $d_{12}$ (which contain an even power of the external field tensors $f_i^{\mu\nu}$) vanish and only $d_6-d_{11}$ are different from zero.

\subsection{Determination of the coefficients}

Having determined the contraction of $\T^{\mu\nu}_5(q_1,q_2)$ with $q^\mu_1$ and $q^\mu_2$ explicitly, we can restrict us to the contraction from the set $k$, $\Lambda^*_1$ and $\Lambda^*_2$ (or alternatively $k$, $\Lambda_1$ and $\Lambda_2$, see \appref{sec:basisvectorsappendix}) if we analyze the general trace given in \eqref{eqn:sfqed_axialvectorcouplingtensortraceA}. This means that, in order to complete the calculation of $\T^{\mu\nu}_5(q_1,q_2)$, we can ignore the action of the derivatives on $kj$ and $kj'$ and also terms in the trace which are e.g. proportional to $\Ftilde^{\mu\nu}$, $\Ftilde^{*\mu\nu}$, $\Ftilde^{2\mu\nu}$, $(\Ftilde\Ftilde^*)^{\mu\nu}$, $(\Ftilde^*\Ftilde)^{\mu\nu}$, $\Ftilde^{2\mu\rho} v_\rho$, $v_\rho \Ftilde^{2\rho\nu}$, where $v^\mu$ is an arbitrary four-vector. In particular, we see that the terms $a^{\alpha\mu} d_{\alpha}^{\phantom{\alpha}\nu}$ and $b^{\alpha\mu}c_{\alpha}^{\phantom{\alpha}\nu}$ can be ignored and therefore also the action of the derivatives on the term in the exponent which is quadratic in the sources. If the derivatives act on the non-quadratic source-terms in the exponent, we obtain the replacement rules
\begin{gather}
\begin{aligned}
p^\alpha p'^\beta + p^\beta p'^\alpha
&\longrightarrow&
-\frac{2\mu}{\tau} q_2^\alpha q_2^\beta &+ \frac{1}{2\tau^2} \uplambda^\alpha \uplambda^\beta
\\
&& &- \frac{v}{2\tau} (q_2^\alpha \uplambda^\beta + \uplambda^\alpha q_2^\beta),
\\
p^\alpha p'^\beta - p^\beta p'^\alpha
&\longrightarrow&
\frac{1}{2\tau} (q_2^\alpha \uplambda^\beta &- q_2^\beta \uplambda^\alpha).
\end{aligned}
\end{gather}
Finally, we can replace the trace given in \eqref{eqn:sfqed_axialvectorcouplingtensortraceA} by (terms linear in $v$ do not contrubute after the integration)
\begin{multline}
\frac14 \tr \big[ \cdots \big]_5^{\mu\nu} 
\longrightarrow 
\I \bigg\{ \frac{e}{kq} \big[q_2^\mu (\Ftilde^*_y q)^\nu - (\Ftilde^{*}_x q)^\mu q_2^\nu \big]
\\+ \frac{1}{2\tau} \eps^{\mu\nu\rho\sigma} q_{2\rho} \uplambda_\sigma \bigg\}
\end{multline}
[as long as only contractions with $k$ and $\Lambda_i$/$\Lambda^*_i$ are considered, which also means that the anomaly does not contribute, see \eqref{eqn:T5anomalyresult_final}].

\subsection{Final result}

Using the relations given in \appref{sec:basisvectorsappendix} we obtain the following representation for the tensor $\T_5^{\mu\nu}(q_1,q_2)$
\begin{multline}
\label{eqn:axialvectorcouplingtensorfinal}
\T_5^{\mu\nu}(q_1,q_2)
= 
\mathfrak{T}_5^{\mu\nu}(q_1,q_2)
-
\I\pi e^2\, \delta^{(\lminus,\lperp)}(q_1-q_2) 
\\ \times
\, \int_{-1}^{+1} dv \int_0^\infty \frac{d\tau}{\tau} \, 
\int_{-\infty}^{+\infty} dz^\lminus \,
\big[
a_6 \Q_1^\mu \Lambda_1^{*\nu}
+
a_7 \Q_1^\mu \Lambda_2^{*\nu}
\\+
a_8 \Lambda_1^{*\mu} \Q_2^{\nu}
+
a_9 \Lambda_2^{*\mu} \Q_2^{\nu}
+
a_{10} q_1^\mu \Lambda_1^{*\nu}
+
a_{11} q_1^\mu \Lambda_2^{*\nu} 
\big]  e^{i\Phi},
\end{multline}
where the anomalous contribution $\mathfrak{T}^{\mu\nu}_5(q_1,q_2)$ is given in \eqref{eqn:axialvectorcouplingtensorfinal_anomaly} and the coefficients read
\begin{gather}
\begin{gathered}
\begin{aligned}
a_6 &=  \I m  \xi_1 
\Big[W_1 + U_1 \frac{m^2}{q_1^2} \frac{\tau}{\mu} \Big] e^{\I\tau\beta},\\
a_7 &=  \I m  \xi_2 
\Big[W_2 + U_2 \frac{m^2}{q_1^2} \frac{\tau}{\mu} \Big] e^{\I\tau\beta},
\end{aligned}
\\
\begin{aligned}
a_8    &= - \I m  \xi_1 V_1 e^{\I\tau\beta},&
a_{10} &= \I m \xi_1 U_1 \frac{m^2}{q_1^2} \frac{\tau}{\mu} e^{\I\tau\beta},
\\
a_9    &= - \I m  \xi_2 V_2 e^{\I\tau\beta},&
a_{11} &= \I m \xi_2 U_2 \frac{m^2}{q_1^2} \frac{\tau}{\mu} e^{\I\tau\beta}.
\end{aligned}
\end{gathered}
\end{gather}
Here, the phases are given by
\begin{align}
\label{eqn:polarizationoperator_fieldindependentphaseA}
e^{\I\Phi}
&=
\exp \lcb i \lsb
(q_2^\lplus -q_1^\lplus) z^\lminus 
+ 
\mu q_1q_2 
- 
\tau m^2 \rsb \rcb,
\\
e^{\I\tau\beta} &= \exp \big[ i \tau m^2 \sum_{i=1,2} \xi_i^2 (I_i^2-J_i) \big],
\end{align}
where [see \eqref{eqn:defIJintegrals}]
\begin{gather}
\begin{aligned}
I_i 
&=
\frac12 \int_{-1}^{+1} d\lambda \, \psi_i(kz - \lambda \mu kq),\\
J_i 
&=
\frac12 \int_{-1}^{+1} d\lambda \, \psi^2_i(kz - \lambda \mu kq).
\end{aligned}
\end{gather}
In the preexponent we have introduced the following symbols
\begin{gather}
\label{eqn:UVWdefs}
\begin{gathered}
U_i = \psi_i(kx) - \psi_i(ky),\\
V_i = \psi_i(kx) - I_i,
\quad
W_i = \psi_i(ky) - I_i,
\end{gathered}
\end{gather}
where $kx=kz-\mu kq$, $ky=kz+\mu kq$ and $\mu = \frac14 \tau (1-v^2)$.

Alternatively, the result in \eqref{eqn:axialvectorcouplingtensorfinal} can be written as
\begin{multline}
\label{eqn:axialvectorcouplingtensorfinalB}
\T_5^{\mu\nu}(q_1,q_2)
= 
\mathfrak{T}_5^{\mu\nu}(q_1,q_2)
-
\I\pi e^2\, \delta^{(\lminus,\lperp)}(q_1-q_2)
\\ \times
\, \int_{-1}^{+1} dv \int_0^\infty \frac{d\tau}{\tau} \, 
\int_{-\infty}^{+\infty} dz^\lminus \,
\big[
a'_6 \Q_1^\mu \Lambda_1^{*\nu}
+
a'_7 \Q_1^\mu \Lambda_2^{*\nu}
\\+
a_8 \Lambda_1^{*\mu} \Q_2^{\nu}
+
a_9 \Lambda_2^{*\mu} \Q_2^{\nu}
+
a'_{10} k^\mu \Lambda_1^{*\nu}
+
a'_{11} k^\mu \Lambda_2^{*\nu} 
\big]  e^{i\Phi},
\end{multline}
where
\begin{gather}
\begin{gathered}
\begin{aligned}
a'_6 &= \I m  \xi_1 W_1  e^{i\tau\beta},&
a_8 &= - \I m  \xi_1 V_1 e^{i\tau\beta},
\\
a'_7 &= \I m  \xi_2 W_2  e^{i\tau\beta},&
a_9  &= - \I m  \xi_2 V_2 e^{i\tau\beta},
\end{aligned}
\\
\begin{aligned}
a'_{10} &= \I m \xi_1 U_1 \frac{m^2}{kq} \frac{\tau}{\mu} e^{i\tau\beta},
\\
a'_{11} &= \I m \xi_2 U_2 \frac{m^2}{kq} \frac{\tau}{\mu} e^{i\tau\beta}.
\end{aligned}
\end{gathered}
\end{gather}
The last two terms ($a'_{10}$ and $a'_{11}$) are responsible for the inhomogeneous Ward-Takahashi identity [see \eqref{eqn:T5wardidentity} and \eqref{eqn:qonecontractionwithtensor_massterm}].

\section{Adler-Bell-Jackiw anomaly}
\label{sec:anomaly}

We will show now explicitly that the anomalous contribution $\mathfrak{T}_5^\nu(q_1,q_2)$ [see \eqref{eqn:qonecontractionwithtensor_anomalous}] to the Ward-Takahashi identity for the current-coupling tensor $T^{\mu\nu}_5(q_1,q_2)$ [see \eqref{eqn:T5wardidentity}] is different from zero.

As pointed out in \secref{sec:T5wardidentity}, the formal application of the relations given in \eqref{eqn:IGammacontraction} to \eqref{eqn:qonecontractionwithtensor_anomalous} would prove that $\mathfrak{T}_5^\nu(q_1,q_2)=0$. However, this procedure leads to divergent expressions and a more careful analysis reveals that the obtained result would be incorrect. To determine the anomalous contribution we rewrite \eqref{eqn:qonecontractionwithtensor_anomalous} as
\begin{multline}
\label{eqn:qonecontractionwithtensor_anomalous2}
\mathfrak{T}_5^\nu(q_1,q_2) 
= 
\lim_{\eps\to 0} 4 (-\I e)^2 \int  \frac{d^4p\, d^4p'}{(2\pi)^8} \int d^4x d^4y 
\\
\times (-\I) \int_0^\infty ds \, \Big[\underbrace{\frac{1}{4}\tr [\cdots]_{5A}^\nu e^{\I\tilde{S}_T}}_{t=\eps}
-
\underbrace{\frac{1}{4}\tr[\cdots]_{5B}^\nu e^{\I\tilde{S}_T}}_{s=\eps,t=s} \Big],
\end{multline}
where the phase $\tilde{S}_T$ is defined in \eqref{eqn:polopinitialphasedefinition} and the traces are given by
\begin{subequations}
\label{eqn:qonecontractionwithtensor_anomaloustraces}
\begin{align}
\label{eqn:qonecontractionwithtensor_anomaloustracesA}
\frac{1}{4}\tr [\cdots]_{5A}^\nu 
=& \I G_3 \big[(p_\mu \Ftilde^{*\mu\nu}_x - p_\mu \Ftilde^{*\mu\nu}_y)
\nonumber\\
&+ 
G_1 \,  p^\rho \Ftilde_{x\rho\mu}^* \Ftilde^{\mu\nu}_y
+ 
G_3 \, p^\rho \Ftilde_{x\rho\mu} \Ftilde^{*\mu\nu}_y\big],
\displaybreak[0]\\
\label{eqn:qonecontractionwithtensor_anomaloustracesB}
\frac{1}{4}\tr [\cdots]_{5B}^\nu 
=& \I G_3 \big[(p'_\mu \Ftilde^{*\mu\nu}_x - p'_\mu \Ftilde^{*\mu\nu}_y)
\nonumber\\
&+ 
G_1 \,  p'^\rho \Ftilde_{x\rho\mu}^* \Ftilde^{\mu\nu}_y
- 
G_3 \, p'^\rho \Ftilde_{x\rho\mu} \Ftilde^{*\mu\nu}_y \big],
\end{align}
\end{subequations}
with $G_i = G_i(kp',kp)$. 

Although we need to exponentiate here only one scalar propagator [see \eqref{eqn:qonecontractionwithtensor_anomalous}], we artificially add a second term in the exponent (multiplied by a quantity $\eps$ which will be later sent to zero), in order to recover exactly the same structure as in \eqref{eqn:polopinitialphasedefinition}. Also note that the traces in \eqref{eqn:qonecontractionwithtensor_anomaloustracesA} and \eqref{eqn:qonecontractionwithtensor_anomaloustracesB} can be formally obtained from \eqref{eqn:qonecontractiontrace} by setting $p'^\mu=0$ and $p^\mu=0$, respectively, and by dividing by $2m^2$ and $-2m^2$, respectively. To match the first and the second contribution in \eqref{eqn:qonecontractionwithtensor_anomalous2}, we changed the name of the integration variable $t\to s$ (after the replacement $s\to\eps$) in the second expression. 

In order to determine the first and the second contribution to \eqref{eqn:qonecontractionwithtensor_anomalous2}, respectively, we need to apply the following replacements to \eqref{eqn:T5withpropertimeintegrals} 
\begin{gather}
\label{eqn:T5anomalyvspolopreplacements}
\begin{aligned}
(-\I) \int_0^\infty dt &\rightarrow \one,&
&t \rightarrow \eps,&
\tr [\cdots]_5^{\mu\nu} &\rightarrow \tr [\cdots]_{5A}^\nu,
\\
(-\I) \int_0^\infty ds &\rightarrow \one,&
&\begin{aligned}s &\rightarrow \eps,\\[-5pt] t &\rightarrow s,\end{aligned}&
\tr [\cdots]_5^{\mu\nu} &\rightarrow \tr [\cdots]_{5B}^\nu.
\end{aligned}
\end{gather}
In this way, the final result can then be obtained from \eqref{eqn:T5afterintegralstaken}. 

The replacements given in \eqref{eqn:T5anomalyvspolopreplacements} imply that $\tau = s+t$ and $\mu = \nfrac{st}{(s+t)}$ are mapped to the same quantity in both expressions, but $v = \nfrac{(s-t)}{(s+t)}$ changes its sign
\begin{gather}
\begin{aligned}
\tau(t\to\eps) &= \phantom{-}\tau(s\to\eps,t\to s) = s+\eps,
\\
\mu(t\to\eps) &= \phantom{-}\mu(s\to\eps,t\to s) = \frac{s\eps}{s+\eps},
\\
v(t\to\eps) &= -v(s\to\eps,t\to s) =  \frac{s-\eps}{s+\eps}.
\end{aligned}
\end{gather}

We note that due to the relation $ky = kx + 2\mu kq$ [see Eq. (57) in \cite{meuren_polarization_2013}], the distance (here in phase) between the two vertices tends to zero as $\eps\to0$. A similar regularization procedure for the axial-vector vertex is also commonly used in the calculation of the vacuum amplitude (see e.g. chapter 19 of \cite{peskin_introduction_2008} for a textbook discussion). 

To use \eqref{eqn:T5afterintegralstaken} we have to apply the replacement rules given in \eqref{eqn:T5_ppprimereplacementruleswardidentity}. Firstly, we note that $\uplambda^\mu$ does not contribute, as it would only give a non-vanishing contribution after contraction with the first line of each trace in \eqref{eqn:qonecontractionwithtensor_anomaloustraces} which then cancel pairwise. Correspondingly, we can focus on the contribution due to $q_2^\mu$. As the second and the third line of each trace cancel both pairwise, we focus on the first line. Using the following representation for the delta function
\begin{gather}
\label{eqn:T5warddeltafuncrep}
\lim_{\eps\to0}
\int_0^\infty dx \, \frac{\eps \, g(x)}{(x+\eps)^2} 
=
\lim_{\eps\to0}
\int_0^{\infty} dy \, \frac{g(\eps y)}{(y+1)^2}
=
g(0)
\end{gather}
[$g(x)$ is assumed to be sufficiently regular], we finally obtain the result given in \eqref{eqn:T5anomalyresult_final}.

\section{Special field configurations} 
\label{sec:T5specialfields}

In this section the general expression given in Eqs. (\ref{eqn:axialvectorcouplingtensorfinal}) and (\ref{eqn:axialvectorcouplingtensorfinalB}) is used to derive compact representations for the current-coupling tensor $\T_5^{\mu\nu}(q_1,q_2)$ for three important special field configurations: a constant-crossed field, a relativistically strong, linearly polarized plane-wave background field (quasistatic limit) and a monochromatic, circularly polarized plane-wave background field. When possible, the result is compared with existing representations from the literature.

\subsection{Constant-crossed field}
\label{sec:T5ccfield}

From \eqref{eqn:axialvectorcouplingtensorfinal} we can derive the result for a constant-crossed field, which is characterized by
\begin{gather}
\label{eqn:constantcrossedfield}
\psi_1(\phi) = \phi,
\quad 
\psi_2 = 0,
\end{gather}
(the latter condition corresponds to $\xi_2=0$ and we will write $\xi = \xi_1$ in the following). For a constant-crossed field the field tensor is given by
\begin{gather}
F^{\mu\nu}
= 
f_1^{\mu\nu}
=
f^{\mu\nu}.
\end{gather}
Since $\psi_2 = 0$ the following functions vanish 
\begin{gather}
I_2 = J_2 = U_2 = V_2 = W_2 = 0
\end{gather}
and due to the simple form of $\psi_1$
\begin{gather}
\begin{gathered}
I_1 = kx + \mu kq,
\quad
J_1 = (kx+\mu kq)^2 + \frac13 (\mu kq)^2,\\
U_1 = -2\mu kq,
\quad
V_1 = -\mu kq,
\quad
W_1 = \mu kq.
\end{gathered}
\end{gather}
Finally, we obtain the following explicit expression for the tensor $\T_5^{\mu\nu}(q_1,q_2)$ in a constant-crossed field
\begin{multline}
\label{eqn:axialvectorcouplingtensorsubstractedCCfield}
\T_5^{\mu\nu}(q_1,q_2)
= 
\mathfrak{T}_5^{\mu\nu}(q_1,q_2)
-
2\I\pi^2 e^2\, \delta^{4}(q_1-q_2) \, \times 
\\ \times
\, \int_{-1}^{+1} dv \int_0^\infty \frac{d\tau}{\tau} \, 
\big[
\tilde{b}^{\text{c}}_6 \Q^\mu \Lambda_1^{*\nu}
+
\tilde{b}^{\text{c}}_8 \Lambda_1^{*\mu} \Q^{\nu}
+
\tilde{b}^{\text{c}}_{10} q^\mu \Lambda_1^{*\nu}
\big]  e^{i\Phi_{\text{c}}},
\end{multline}
where the coefficients are given by
\begin{gather}
\begin{aligned}
\tilde{b}^{\text{c}}_6 &=  i \xi m kq  
\Big[\frac{1}{w}   - \frac{2m^2}{q^2} \Big] \tau e^{i\tau\beta_{\text{c}}},\\
\tilde{b}^{\text{c}}_8 &=  i \xi m kq \, \frac{1}{w} \tau e^{i\tau\beta_{\text{c}}},\\
\tilde{b}^{\text{c}}_{10} &= i \xi m kq \, (-\nfrac{2m^2}{q^2})  \tau e^{i\tau\beta_{\text{c}}},
\end{aligned}
\end{gather}
the phases read
\begin{gather}
\begin{aligned}
i\Phi_{\text{c}} &=  -i\tau a,& a &=  m^2 \lsb 1-\frac{1}{4}(1-v^2) \frac{q^2}{m^2}\rsb,\\
i\tau\beta_{\text{c}} &= - \frac{i}{3} \tau^3 b,& b &= m^6 \chi^2 \lsb \frac14(1-v^2)\rsb^2
\end{aligned}
\end{gather}
and the anomaly $\mathfrak{T}_5^{\mu\nu}(q_1,q_2)$ [see \eqref{eqn:axialvectorcouplingtensorfinal_anomaly}] becomes
\begin{multline}
\label{eqn:axialvectorcouplingtensorsubstractedCCfieldfinal_anomaly}
\mathfrak{T}^{\mu\nu}_5(q_1,q_2)
= 
\I(2\pi)^4 \delta^{4}(q_1-q_2)  
\\\times\,
\lb-\frac{e^3}{8\pi^2 m^2}\rb\,
4\frac{m^2}{kq} \lsb k^\mu (qF^{*})^\nu + (qF^{*})^\mu k^\nu \rsb.
\end{multline}
Above, we introduced the quantum-nonlinearity parameter
\begin{gather}
\label{eqn:chidefinition}
\chi 
= 
-\frac{e\sqrt{qF^2q}}{m^3}
= 
\xi \frac{\sqrt{(kq)^2}}{m^2}.
\end{gather}
Due to the overall momentum-conserving delta function we define
\begin{gather}
q^\mu = q_1^\mu = q_2^\mu,
\quad
\Q^\mu = \Q_1^\mu = \Q_2^\mu = \frac{k^\mu q^2 - q^\mu kq}{kq}.
\end{gather}
Using the relation
\begin{multline}
\tilde{b}^{\text{c}}_6 \Q^\mu \Lambda^{*\nu}_1
+
\tilde{b}^{\text{c}}_{10} q^\mu \Lambda^{*\nu}_1
=
\frac{e}{\xi m kq} (\tilde{b}^{\text{c}}_6 - \tilde{b}^{\text{c}}_{10}) \, q^\mu (F^*q)^\nu 
\\-
\frac{e}{\xi m kq} \frac{q^2}{kq}  \tilde{b}^{\text{c}}_6 \, k^\mu (F^*q)^\nu
\end{multline}
we can rewrite \eqref{eqn:axialvectorcouplingtensorsubstractedCCfield} as
\begin{multline}
\label{eqn:axialvectorcouplingtensorsubstractedCCfieldfinal}
\T_5^{\mu\nu}(q_1,q_2)
=
\mathfrak{T}_5^{\mu\nu}(q_1,q_2)
+ 
\I(2\pi)^4 \delta^{4}(q_1-q_2) 
\\ \times
\big[
\tilde{\tau}_{1} \Q^\mu (F^*q)^\nu + \tilde{\tau}_{2} k^\mu (F^*q)^\nu + \tilde{\tau}_{1} (F^*q)^\mu \Q^\nu
\big],
\end{multline}
where
\begin{gather}
\begin{aligned}
\tilde{\tau}_{1} 
&=
+ \frac{e^3}{8\pi^2m^2} \int_{-1}^{+1} dv \, \frac{1}{w} \Big(\frac{w}{\chi}\Big)^{\nicefrac23} f(\rho),
\\
\tilde{\tau}_{2} 
&=
- \frac{e^3}{8\pi^2m^2} \int_{-1}^{+1} dv \, 2\frac{m^2}{kq}  \Big(\frac{w}{\chi}\Big)^{\nicefrac23} f(\rho)
\end{aligned}
\end{gather}
[$\tfrac{1}{w} = \tfrac14(1-v^2)$, $\rho = \big(\tfrac{w}{\chi}\big)^{\nicefrac23}(1- \tfrac{q^2}{m^2} \tfrac{1}{w})$] and the anomaly is given in \eqref{eqn:axialvectorcouplingtensorsubstractedCCfieldfinal_anomaly}. Furthermore, the Ritus functions are defined by \cite{ritus_radiative_1972,ritus_1985}
\begin{subequations}
\label{eqn:polop_ritusfunctions}
\begin{multline}
f(x)
= 
\I \int_0^\infty dt \exp\big[-\I\big( t x + \nfrac{t^3}{3}\big)\big] 
\\=
\pi \Gi(x) + \I \pi \Ai(x),
\end{multline}
\begin{gather}
f_1(x)
=
\int_0^\infty \frac{dt}{t} \exp\lb-\I t x \rb \lsb \exp\big(- \I \nfrac{t^3}{3} \big) -1\rsb,
\end{gather}
\end{subequations}
where $\Ai$ and $\Gi$ are the Airy and Scorer function, respectively \cite{olver_nist_2010}. Note that in Ritus' work the normalization of the Airy function is different and also changes [see \cite{ritus_1985}, App.\,C and \cite{nikishov_quantum_1964}, Eq.\,(B5)].

Since all nonvanishing functions are even in $v$, it is possible to apply the following change of variables
\begin{gather}
\int_{-1}^{+1} dv = 2 \int_{0}^{1} dv = \int_{4}^{\infty} dw \, \frac{4}{w\sqrt{w(w-4)}}.
\end{gather}

The final result given in \eqref{eqn:axialvectorcouplingtensorsubstractedCCfieldfinal} coincides with the one given in Eq.\,(4.24) of \cite{schubert_vacuum_2000-1}, apart from the anomalous contribution in the vector index [see \eqref{eqn:T5wardidentity}], which automatically drops out by performing the calculations within the wordline formalism as in \cite{schubert_vacuum_2000-1}. If evaluated on the mass shell (i.e. for $q^2=0$), it also agrees with Eq.\,(15) in \cite{shaisultanov_2000}.

\subsection{Linear polarization}
\label{sec:T5qs}

We consider now a linearly polarized plane-wave field
\begin{gather}
\label{eqn:linearpolarizationdef}
\psi_1(\phi) = \psi(\phi), 
\quad
\psi_2= 0 
\end{gather}
($\xi = \xi_1$, $f^{\mu\nu} = f_1^{\mu\nu}$) in the quasistatic limit defined by $\xi\to\infty$ while [see \eqref{eqn:chidefinition}]
\begin{gather}
\chi 
= 
-\frac{e\sqrt{qf^2q}}{m^3}
= 
\xi \frac{\sqrt{(kq)^2}}{m^2}
\end{gather}
is kept constant. In the optical regime (photon energy $\omega_0 \sim \unit[1]{eV}$) the condition $\chi \gtrsim 1$ usually requires $\xi \gg 1$, which means that the quasistatic limit is in general sufficient to analyze strong-field experiments with optical lasers (it neglects the recollision contribution considered in \cite{meuren_high-energy_2015}, though).  

For a linearly polarized background field we obtain
\begin{gather}
I_2 = J_2 = U_2 = V_2 = W_2 = 0
\end{gather}
and using the relation $\abs{kq} = \nfrac{m^2 \chi}{\xi}$ it is sufficient to consider the leading-order contribution to the following quantities
\begin{gather}
\label{eqn:qclimit_IJUVW}
\begin{aligned}
I^2_1 - J_1 &= -(\nfrac13) (\mu kq)^2 \big[\psi'(kz)\big]^2 + \mc{O}(\mu kq)^3,\\
U_1 &= - 2\mu kq \psi'(kz) + \mc{O}(\mu kq)^2,\\
V_1 &= - \mu kq \psi'(kz) + \mc{O}(\mu kq)^2,\\
W_1 &= + \mu kq \psi'(kz) + \mc{O}(\mu kq)^2.
\end{aligned}
\end{gather}
If we insert these relations into \eqref{eqn:axialvectorcouplingtensorfinalB}, the remaining calculation is very similar to the one for a constant-crossed field (see \secref{sec:T5ccfield}), the essential change is the replacement $\chi \to \chi(kz) = \chi \abs{\psi'(kz)}$. The final result is given by
\begin{multline}
\label{eqn:axialvectorcouplingtensorqcapproxfinal}
\T_5^{\mu\nu}(q_1,q_2)
=
\mathfrak{T}_5^{\mu\nu}(q_1,q_2)
+ 
\I(2\pi)^4 \delta^{(\lminus,\lperp)}(q_1-q_2) \, \times \\ \times \,
\frac{1}{2\pi}
\int_{-\infty}^{+\infty} dz^\lminus e^{i(q_2^\lplus - q_1^\lplus) z^\lminus}
\, \psi'(kz)  \big[ \tau'_1 \Q_1^\mu (f^*q)^\nu \\+ \tau_1' (f^*q)^\mu \Q_2^\nu + \tau'_2 k^\mu (f^*q)^\nu \big]
\end{multline}
where 
\begin{gather}
\begin{aligned}
\tau'_1 &= +\frac{e^3}{8\pi^2 m^2} \int_{-1}^{+1} dv \, \frac{1}{w} \lsb \frac{w}{\abs{\chi(kz)}} \rsb^{\nicefrac23} f(\rho),
\\
\tau'_2 &= -\frac{e^3}{8\pi^2 m^2} \int_{-1}^{+1} dv \, \frac{2m^2}{kq} \lsb \frac{w}{\abs{\chi(kz)}} \rsb^{\nicefrac23} f(\rho)
\end{aligned}
\end{gather}
and $\frac{1}{w} = \frac14(1-v^2)$, $\rho = \big[\frac{w}{\abs{\chi(kz)}}\big]^{\nicefrac23}(1- \frac{q_1 q_2}{m^2} \frac{1}{w})$. Furthermore, the anomaly reads [see \eqref{eqn:axialvectorcouplingtensorfinal_anomaly}]
\begin{multline}
\mathfrak{T}^{\mu\nu}_5(q_1,q_2)
= \I (2\pi)^4 \delta^{(\lminus,\lperp)}(q_1-q_2) 
\\
\begin{aligned}
&\times
\, \frac{1}{2\pi} \int_{-\infty}^{+\infty} dz^\lminus \, e^{\I(q_2^\lplus-q_1^\lplus) z^\lminus}  \, \psi'(kz) 
\\
&\times \, \lb -\frac{e^3}{8\pi^2 m^2} \rb 4 \frac{m^2}{kq} \lsb k^\mu (qf^{*})^\nu + (qf^{*})^\mu k^\nu \rsb.
\end{aligned}
\end{multline}
Note that for $\psi'(\phi) = 1$ the result given in \eqref{eqn:axialvectorcouplingtensorqcapproxfinal} coincides (as required) with the one for a constant-crossed field [see \eqref{eqn:axialvectorcouplingtensorsubstractedCCfieldfinal}].

\subsection{Circular polarization}
\label{sec:T5circular}

Also for a circularly polarized, monochromatic background field
\begin{gather}
\psi_1(\phi) = \Re{e^{i\phi}},
\quad
\psi_2(\phi) = \Im{e^{i\phi}},
\quad
\xi_1=\xi_2=\xi
\end{gather}
the result given in \eqref{eqn:axialvectorcouplingtensorfinalB} simplifies considerably and we obtain
\begin{multline}
\label{eqn:axialvectorcouplingcircularpolarizationA}
\T_5^{\mu\nu}(q_1,q_2)
= 
\mathfrak{T}_5^{\mu\nu}(q_1,q_2)
-
\I\pi e^2\, \delta^{(\lminus,\lperp)}(q_1-q_2) \, \times 
\\ \times
\, \int_{-1}^{+1} dv \int_0^\infty \frac{d\tau}{\tau} \, 
\int_{-\infty}^{+\infty} dz^\lminus \,
\big[
a^+_{1} \Q_1^\mu \widetilde{\Lambda}_+^{\nu}
+
a^-_{1} \Q_1^\mu \widetilde{\Lambda}_-^{\nu}
\\+
a^+_{2} \widetilde{\Lambda}_+^{\mu} \Q_2^{\nu}
+
a^-_{2} \widetilde{\Lambda}_-^{\mu} \Q_2^{\nu}
+
a^+_{3} k^\mu \widetilde{\Lambda}_+^{\nu}
+
a^-_{3} k^\mu \widetilde{\Lambda}_-^{\nu} 
\big]  e^{i\Phi},
\end{multline}
where the anomaly is given in \eqref{eqn:axialvectorcouplingtensorfinal_anomaly} and
\begin{gather}
\begin{aligned}
a_1^+ &= \frac12 (a_6'-i a_7') 
= 
\frac12 i m \xi (W_1-i W_2)  e^{i\tau\beta},
\\
a_1^- &= \frac12 (a_6'+i a_7') 
=
\frac12 i m \xi (W_1+i W_2)  e^{i\tau\beta},
\\
a_2^+ &= \frac12 (a_8-i a_9) 
= 
-\frac12 i m \xi (V_1-i V_2)  e^{i\tau\beta},
\\
a_2^- &= \frac12 (a_8+i a_9) 
=
-\frac12 i m \xi (V_1+i V_2)  e^{i\tau\beta},
\\
a_3^+ &= \frac12 (a'_{10}-i a'_{11}) 
= 
\frac12 i m \xi \frac{\tau m^2}{\mu kq} (U_1-i U_2)e^{i\tau\beta},
\\
a_3^- &= \frac12 (a'_{10}+i a'_{11}) 
=
\frac12 i m \xi \frac{\tau m^2}{\mu kq}  (U_1+i U_2) e^{i\tau\beta},
\end{aligned}
\end{gather}
\begin{gather}
\begin{aligned}
i\tau\beta 
&= 
i \tau m^2 \xi^2 \lsb \sinc^2(\mu kq) -1 \rsb,\\
i \Phi
&=
i \lsb (q_2^\lplus -q_1^\lplus) z^\lminus  +  \mu q_1q_2 -  \tau m^2 \rsb
\end{aligned}
\end{gather}
and
\begin{gather}
\widetilde{\Lambda}^{\mu}_{\pm} = \Lambda^{*\mu}_1 \pm \I \Lambda^{*\mu}_2
\end{gather}
(the star is part of the symbol, both $\Lambda^{*\mu}_1$ and $\Lambda^{*\mu}_2$ are real four-vectors). Furthermore,
\begin{gather}
\begin{aligned}
W_1+i W_2 &= -A,&
W_1-i W_2 &= -A^*,\\
V_1+i V_2 &= -B,&
V_1-i V_2 &= -B^*,\\
U_1+i U_2 &= -C,&
U_1-i U_2 &= -C^*\\
\end{aligned}
\end{gather}
where
\begin{gather}
\begin{aligned}
A &= e^{ikz} \lsb \sinc(\mu kq) - \cos(\mu kq) - i \sin(\mu kq) \rsb,\\
B &= e^{ikz} \lsb \sinc(\mu kq) - \cos(\mu kq) + i \sin(\mu kq) \rsb,\\
C &= e^{ikz} \, 2i \sin(\mu kq)
\end{aligned}
\end{gather}
and therefore
\begin{gather}
\begin{aligned}
-W_1 &= I_{1} - \psi_{1}(kz+\mu kq)
=
\Re A,\\
-W_2 &= I_{2} - \psi_{2}(kz+\mu kq)
=
\Im A,\\
-V_1 &= I_{1} - \psi_{1}(kz-\mu kq)
=
\Re B,\\
-V_2 &= I_{2} - \psi_{2}(kz-\mu kq)
=
\Im B,\\
-U_1 &=  \psi_{1}(kz+\mu kq)-\psi_{1}(kz-\mu kq)
=
\Re C,\\
-U_2 &=  \psi_{2}(kz+\mu kq)-\psi_{2}(kz-\mu kq)
=
\Im C.
\end{aligned}
\end{gather}
We can now take the integral in $dz^-$ and obtain
\begin{multline}
\label{eqn:axialvectorcouplingcircularpolarizationB}
\T_5^{\mu\nu}(q_1,q_2)
=
\mathfrak{T}_5^{\mu\nu}(q_1,q_2) 
-
\I(2\pi)^4 \frac{e^2}{8\pi^2} \, \times \\ \times \,  \int_{-1}^{+1} dv 
\int_0^\infty \frac{d\tau}{\tau} \,
\big[ T_{5+}^{\mu\nu} \, \delta(q_1-q_2+k) \\+ T_{5-}^{\mu\nu} \, \delta(q_1-q_2-k)  \big]
e^{i\Phi_{\mathrm{cp}}},
\end{multline}
where
\begin{gather}
i\Phi_{\text{cp}}
=
- i\tau m^2 \big\{1 + \xi^2 [1 - \sinc^2(\mu kq)] \big\} + i \mu q_1q_2,
\end{gather}
\begin{gather}
\label{eqn:T5pmcircpol}
T_{5\pm}^{\mu\nu} 
= 
(\lambda^\pm_{1} \Q_1^\mu + \lambda^\pm_{3} k^\mu) \widetilde{\Lambda}_\pm^{\nu}
+
\lambda^\pm_{2} \widetilde{\Lambda}_\pm^{\mu} \Q_2^{\nu}
\end{gather}
and
\begin{gather}
\label{eqn:T5lambdaicricpol}
\begin{aligned}
\lambda^\pm_{1} &= - \frac12 i m \xi \lsb \sinc(\mu kq) - \cos(\mu kq) \pm i \sin(\mu kq) \rsb,
\\
\lambda^\pm_{2} &= +\frac12 i m \xi  \lsb \sinc(\mu kq) - \cos(\mu kq) \mp i \sin(\mu kq) \rsb,
\\
\lambda^\pm_{3} &= \mp \, m \xi \, \tau  m^2 \sinc(\mu kq).
\end{aligned}
\end{gather}
Correspondingly, the result is in agreement with the one obtained in \cite{gvozdev_radiative_1993,gvozdev_electromagnetic_1994,gvozdev_radiative_1996}.

\section{Conclusion}
\label{sec:conclusion}

In the present paper the axial-vector--vector current-coupling tensor $\T_5^{\mu\nu}(q_1,q_2)$ [see \figref{fig:T5vsT} and \eqref{eqn:axialvectorcouplingtensor}] has been considered for the first time in the presence of a general plane-wave background field (arbitrary polarization and pulse shape). The (pseudo-)tensor $\T_5^{\mu\nu}(q_1,q_2)$ appears in the calculation of various electroweak processes inside strong laser fields like photon emission by neutrinos [see \figref{fig:neutrinoprocesses} and \eqref{eqn:neutrinophotonemission_matrixelementZboson}]. We derived a triple-integral representation for $\T_5^{\mu\nu}(q_1,q_2)$ [see Eqs. (\ref{eqn:axialvectorcouplingtensorfinal}) and (\ref{eqn:axialvectorcouplingtensorfinalB})], which can even be converted into a double-integral representation (see \cite{meuren_polarization-operator_2015} for details). In particular, the anomalous contribution to the Ward-Takahashi identity [see \eqref{eqn:T5wardidentity}] due to the Adler-Bell-Jackiw (ABJ) anomaly associated with the axial-vector current has been calculated explicitly [see \figref{fig:anomaly} and \eqref{eqn:T5anomalyresult_final}]. Finally, we specialized the obtained general expression to three important types of background plane waves and confirmed agreement with the corresponding results available in the literature: a constant-crossed field [see \eqref{eqn:axialvectorcouplingtensorsubstractedCCfieldfinal}], a relativistically strong, linearly polarized plane-wave background field [see \eqref{eqn:axialvectorcouplingtensorqcapproxfinal}] and a monochromatic, circularly polarized plane-wave background field [see \eqref{eqn:axialvectorcouplingcircularpolarizationB}].

\begin{acknowledgments}
S.M. is grateful to the Studienstiftung des deutschen Volkes for financial support.
\end{acknowledgments}

\appendix

\section{Dressed vertex}
\label{sec:dressedvertexappendix}

A strong plane-wave background field can be taken into account exactly by using dressed states for the charged particles. In momentum space this implies that the free vertex must be replaced by the so-called dressed vertex (see e.g. \cite{meuren_polarization_2013,mitter_quantum_1975} for more details). Using the Ritus $E_p$ matrices \cite{ritus_1985,mitter_quantum_1975}
\begin{gather}
\begin{aligned}
E_{p,x} &= \lsb\one + \frac{e\s{k}\s{A}(kx)}{2\, kp}\rsb \, e^{iS_p(x)},
\\
\bar{E}_{p,x}  &=  \lsb\one + \frac{e\s{A}(kx)\s{k}}{2\, kp}\rsb \, e^{-iS_p(x)}
\end{aligned}
\end{gather}
which contain the Volkov action
\begin{gather}
S_p(x) 
=
- px - \int_{-\infty}^{kx} \lsb \frac{e\, pA(\phi')}{\, kp} - \frac{e^2A^2(\phi')}{2\, kp} \rsb d\phi',
\end{gather}
we define the dressed vector ($\Gamma^\rho$) and scalar ($I$) vertices by
\begin{align}
\label{eqn:sfqed_dressedvertex}
\Gamma^\rho(p',q,p) 
&= 
-\I e  \int d^4x \, e^{-\I qx}\, \bar{E}_{p',x} \gamma^\rho E_{p,x},
\\
\label{eqn:sfqed_Ivertexdefinition}
I(p',q,p) 
&=
-\I e  \int d^4x \, e^{-\I qx}\, \bar{E}_{p',x} E_{p,x}.
\end{align}
They can be decomposed in the following way \cite{meuren_polarization_2013}
\begin{subequations}
\begin{multline}
\Gamma^\rho(p',q,p) 
= 
-\I e  \int d^4x \,\big[ \gamma_\mu G^{\mu\rho}(kp',kp;kx) 
\\ \I \gamma_\mu \gamma^5 G^{\mu\rho}_5(kp',kp;kx) \big]  e^{\I S_\Gamma},
\end{multline}
\begin{gather}
I(p',q,p) 
= 
-ie \int d^4x \, \lsb \one + \frac{G_3}{2} \sigma^{\alpha\beta} \Ftilde_{\alpha\beta}(kx) \rsb e^{i S_\Gamma},
\end{gather}
\end{subequations}
where we introduced the phase
\begin{multline}
\label{eqn:sfqed_dressedvertexphase}
S_\Gamma = S_\Gamma(p',q,p;x)
=
-S_{p'}(x) -qx  + S_p(x)
\\=
(p'-q-p)x
+
\int_{-\infty}^{kx} d\phi'
\,  \Big[ \frac{e p_\mu p'_\nu \Ftilde^{\mu\nu}(\phi')}{(kp)(kp')} 
\\+ \frac{e^2(kp-kp')}{2(kp)^2(kp')^2} p_\mu p'_\nu \Ftilde^{2\mu\nu}(\phi') \Big]
\end{multline}
and the following tensors
\begin{gather}
\label{eqn:dressedvertexGtensorsdefinition}
\begin{aligned}
G^{\mu\rho}(kp',kp;kx)
&=
g^{\mu\rho} 
+ 
G_1  \Ftilde^{\mu\rho}_x
+ 
G_2 \Ftilde^{2\mu\rho}_x,\\
G_{5}^{\mu\rho}(kp',kp;kx)
&=
G_3 \Ftilde^{*\mu\rho}_x,
\end{aligned}
\end{gather}
\begin{gather}
\label{eqn:Gidefs}
\begin{aligned}
G_1 &= G_1(kp',kp) =  -e\, \frac{kp + kp'}{2 kp \, kp'},
\\
G_2 &= G_2(kp',kp) = \phantom{-e\,}\frac{e^2}{2kp\, kp'},
\\
G_3 &= G_3(kp',kp) = -e\, \frac{kp - kp'}{2 kp \, kp'}.
\end{aligned}
\end{gather}

Finally, we note that the dressed vector and scalar vertices are related by \cite{mitter_quantum_1975}
\begin{multline}
\label{eqn:algebraicidentitydressedvertex}
q_\rho \Gamma^\rho(p',q,p)
\\=
(\s{p}' - m) I(p',q,p) - I(p',q,p) (\s{p} - m)
\end{multline}
and we obtain
\begin{gather}
\label{eqn:IGammacontraction}
\begin{aligned}
\int \frac{d^4p''}{(2\pi)^4}
I(p,q',p'') \Gamma^\mu(p'',q,p')
&=
-\I e \Gamma^\mu (p,q+q',p'),\\
\int \frac{d^4p''}{(2\pi)^4}
\Gamma^\mu(p,q,p'') I(p'',q',p') 
&=
-\I e \Gamma^\mu (p,q+q',p').
\end{aligned}
\end{gather}

\section{Gamma algebra}
\label{sec:gammaalgebraappendix}

Any $4\times 4$ matrix $\Gamma$ in spinor space can be decomposed into five fundamental constituents \cite{meuren_polarization_2013,leader_spin_2001}
\begin{gather}
\label{eqn:decompositionofspinormatrixinfundamentalterms}
\Gamma = c_\one \one + c_5 \gamma^5 + c_\mu \gamma^\mu + c_{5\mu} i\gamma^\mu \gamma^5 + c_{\mu\nu} i \sigma^{\mu\nu},
\end{gather}
where
\begin{gather}
\label{eqn:gammamatrixcoefficients}
\begin{gathered}
c_\one =  \frac{1}{4} \tr \one \Gamma,
\quad
c_5 =  \frac{1}{4} \tr \gamma^5 \Gamma,
\quad
c_\mu =  \frac{1}{4} \tr \gamma_\mu \Gamma,\\
c_{5\mu} =  \frac{1}{4} \tr i \gamma_{\mu} \gamma^5 \Gamma,
\quad
c_{\mu\nu} = \frac{1}{8} \tr i \sigma_{\mu\nu} \Gamma.
\end{gathered}
\end{gather}
Instead of the vector and the axial-vector current one can also use the left- and the right-handed current
\begin{gather}
c_\mu \gamma^\mu + c_{5\mu} i\gamma^\mu \gamma^5
=
l_\mu \gamma^\mu P_L + r_\mu \gamma^\mu P_R,
\end{gather}
where the chirality projectors for the left and the right-handed component are given by
\begin{gather}
P_L = \frac12 \lb \one + \gamma^5\rb,
\quad 
P_R = \frac12 \lb \one - \gamma^5\rb
\end{gather}
(note that we define $\gamma^5 = -i\gamma^0\gamma^1\gamma^2\gamma^3$ as in \cite{landau_quantum_1981}). The coefficients $c_\mu$, $c_{5\mu}$ and $l_\mu$, $r_\mu$ are related via
\begin{gather}
l_\mu = c_\mu + \I c_{5\mu},
\quad
r_\mu = c_\mu - \I c_{5\mu}
\end{gather}
and
\begin{gather}
c_\mu = \frac12 (l_\mu + r_\mu),
\quad
c_{5\mu} = \frac{\I}{2} (r_\mu - l_\mu).
\end{gather}
Therefore, the coefficients for the left- and the right-handed current can be determined from the following traces
\begin{gather}
l_\mu = \frac12 \tr P_L \gamma_\mu  \Gamma,
\quad
r_\mu = \frac12 \tr P_R \gamma_\mu  \Gamma.
\end{gather}

Finally, we note the following contraction identities
\begin{gather}
\label{eqn:gammacontractionidentitiesfundamentalterms}
\begin{gathered}
\begin{aligned}
\gamma^\rho \one     \gamma_\rho &= \phantom{+}4,&
\gamma^\rho \gamma^5 \gamma_\rho &= -4 \gamma^5,
\\
\gamma^\rho \gamma^\mu \gamma_\rho &= -2 \gamma^\mu,&
\gamma^\rho (\I\gamma^\mu \gamma^5) \gamma_\rho &= \phantom{+} 2 (\I\gamma^\mu\gamma^5),
\\
\end{aligned}
\\
\gamma^\rho (\I\sigma^{\mu\nu}) \gamma_\rho = 0.
\end{gathered}
\end{gather}

\section{Summary of important relations}
\label{sec:basisvectorsappendix}

To obtain a simple structure we expanded the polarization tensor $\T^{\mu\nu}(q_1,q_2)$ using the two complete sets $q_1^\mu$, $\Q_1^\mu$, $\Lambda_1^\mu$, $\Lambda_2^\mu$ and $q_2^\nu$, $\Q_2^\nu$, $\Lambda_1^\nu$, $\Lambda_2^\nu$ \cite{baier_interaction_1975}, where
\begin{gather}
\label{eqn:LambdaandQvectors}
\begin{gathered}
\Lambda_1^\mu = \frac{f_1^{\mu\nu} q_\nu}{kq \sqrt{-a_1^2}},
\quad
\Lambda_2^\mu = \frac{f_2^{\mu\nu} q_\nu}{kq \sqrt{-a_2^2}},
\\
\Q_1^\mu = \frac{k^\mu q_1^2 - q_1^\mu kq}{kq},
\quad
\Q_2^\mu = \frac{k^\mu q_2^2 - q_2^\mu kq}{kq}.
\end{gathered}
\end{gather}
They have the following properties
\begin{gather}
\label{eqn:LambdaQqproperties}
\begin{gathered}
\Lambda_i \Lambda_j = -\delta_{ij},
\quad
k\Lambda_i = q_i\Lambda_j = \Q_i \Lambda_j = 0,\
\\
\Q_1^2 = -q_1^2, 
\quad
\Q_2^2 = -q_2^2,
\quad
q_i \Q_i= 0.
\end{gathered}
\end{gather}

Since $\T_5^{\mu\nu}(q_1,q_2)$ includes $\Ftilde^{*\mu\nu}$ [while $\T^{\mu\nu}(q_1,q_2)$ contains $\Ftilde^{\mu\nu}$] it is more natural to expand $\T_5^{\mu\nu}(q_1,q_2)$ using the two complete sets $q_1^\mu$, $\Q_1^\mu$, $\Lambda^{*\mu}_1$, $\Lambda^{*\mu}_2$ and $q_2^\nu$, $\Q_2^\nu$, $\Lambda^{*\nu}_1$, $\Lambda^{*\nu}_2$ where 
\begin{gather}
\label{eqn:Lambdastarvectors}
\Lambda_1^{*\mu} = \frac{f_1^{*\mu\nu} q_\nu}{kq \sqrt{-a_1^2}},
\quad
\Lambda_2^{*\mu} = \frac{f_2^{*\mu\nu} q_\nu}{kq \sqrt{-a_2^2}}.
\end{gather}
The (pseudo) four-vectors $\Lambda^{*\mu}_i$ have similar properties as the four-vectors $\Lambda^\mu_i$ [compare with \eqref{eqn:LambdaQqproperties}]
\begin{gather}
\begin{aligned}
\Lambda^*_i \Lambda^*_j &= -\delta_{ij},&
k\Lambda^*_i &= q_i\Lambda^*_j = \Q_i\Lambda^*_j = 0.
\end{aligned}
\end{gather}
Thus, $\Lambda^\mu_i$ and $\Lambda^{*\mu}_i$ span the same subspace and we obtain the following scalar products
\begin{gather}
\Lambda_i \Lambda^*_i =  0,
\quad
\Lambda_1^* \Lambda_2 = \frac{q\Lambda_5}{kq},
\quad
\Lambda_1 \Lambda^*_2 = -\frac{q\Lambda_5}{kq},
\end{gather}
where we defined the pseudo four-vector
\begin{gather}
\Lambda_5^\mu 
= 
\frac{\eps^{\mu\nu\rho\sigma} k_\nu a_{1\rho} a_{2\sigma}}{\sqrt{-a_1^2}\sqrt{-a_2^2}}
=
\frac{q\Lambda_5}{kq} k^\mu,
\end{gather}
which obeys
\begin{gather}
\Lambda_5^2 = \Lambda_5k = \Lambda_5\Lambda_1 = \Lambda_5\Lambda_2 = 0,
\quad
(q\Lambda_5)^2 = (kq)^2
\end{gather}
[beside being a pseudo four-vector $\Lambda_5^\mu$ is proportional to $k^\mu$]. Thus, we obtain the identities 
\begin{gather}
\Lambda_1^\mu = \frac{kq}{q\Lambda_5} \Lambda_2^{*\mu},
\quad
\Lambda_2^\mu = - \frac{kq}{q\Lambda_5} \Lambda_1^{*\mu}.
\end{gather}

We note the following relations
\begin{gather}
f^{\mu}_{i\,\rho} f_{j}^{\rho\nu} 
= 
-\delta_{ij} a_i^2\, k^\mu k^\nu,
\quad
f^{*\mu}_{i\,\phantom{*}\rho} f_{j}^{*\rho\nu} 
= 
-\delta_{ij} a_i^2\, k^\mu k^\nu,
\displaybreak[0]\\
\begin{aligned}
f_i^{*\mu\rho} f_{i\rho\nu}
&=
f_i^{\mu\rho} f^*_{i\rho\nu}
=
0,\\
f_1^{\mu\rho} f^{*\,\,\nu}_{2\rho} 
&=
\phantom{-}\sqrt{\smash[b]{-a_1^2}}\sqrt{\smash[b]{-a_2^2}}\, k^\mu \Lambda^\nu_5,
\\
f_1^{*\mu\rho} f^{\phantom{2\rho}\nu}_{2\rho} 
&=
-\sqrt{\smash[b]{-a_1^2}}\sqrt{\smash[b]{-a_2^2}}\, \Lambda^\mu_5 k^\nu,
\\
f_2^{\mu\rho} f^{*\,\,\nu}_{1\rho} 
&=
-\sqrt{\smash[b]{-a_1^2}}\sqrt{\smash[b]{-a_2^2}}\, k^\mu \Lambda^\nu_5,
\\
f_2^{*\mu\rho} f^{\phantom{1\rho}\nu}_{1\rho} 
&=
\phantom{-}\sqrt{\smash[b]{-a_1^2}}\sqrt{\smash[b]{-a_2^2}}\, \Lambda^\mu_5 k^\nu
\end{aligned}
\end{gather}
which imply
\begin{gather}
\begin{aligned}
\Ftilde^{\mu\nu}_x \Lambda_{i\nu}
&=
- \frac{m}{e} k^\mu \xi_i \psi_i(kx),
\\
\Ftilde^{*\mu\nu}_x \Lambda_{1\nu}
&=
-\frac{m}{e} \Lambda^\mu_5 \xi_2 \psi_2(kx),
\\
\Ftilde^{*\mu\nu}_x \Lambda_{2\nu}
&=
\phantom{-}\frac{m}{e} \Lambda^\mu_5 \xi_1 \psi_1(kx)
\end{aligned}
\end{gather}
and
\begin{gather}
\begin{aligned}
\Ftilde^{*\mu\nu}_x \Lambda^*_{i\nu}
&=
- \frac{m}{e} k^\mu \xi_i \psi_i(kx),
\\
\Ftilde^{\mu\nu}_x \Lambda^*_{1\nu}
&=
\phantom{-}\frac{m}{e} \frac{q\Lambda_5}{kq} k^\mu \xi_2 \psi_2(kx),
\\
\Ftilde^{\mu\nu}_x \Lambda^*_{2\nu}
&=
-\frac{m}{e} \frac{q\Lambda_5}{kq} k^\mu \xi_1 \psi_1(kx).
\end{aligned}
\end{gather}
Correspondingly, we obtain the canonical choices $\Ftilde \leftrightarrow \Lambda_{i}$ and $\Ftilde^* \leftrightarrow \Lambda^*_{i}$ (especially for linearly polarized background fields the appearance of $\Lambda_5^\mu$ is unnatural, since its definition involves both $a_1^\mu$ and $a_2^\mu$).

We also note that
\begin{gather}
\eps^{\mu\nu\rho\sigma} k_\rho q_\sigma
=
- q\Lambda_5 \, (\Lambda_1^\mu \Lambda_2^\nu - \Lambda_2^\mu \Lambda_1^\nu)
\end{gather}
and
\begin{gather}
\begin{aligned}
\Ftilde_x^{\mu\rho} \Ftilde_{y\rho\nu}
&=
\frac{m^2}{e^2} k^\mu k_\nu  \sum_{i=1,2}  \xi_i^2 \psi_i(kx) \psi_i(ky),
\\
\Ftilde_x^{*\mu\rho} \Ftilde_{y \rho\nu}
&=
- \frac{m^2}{e^2} \xi_1\xi_2 \, \Lambda_5^\mu  k_\nu 
\,\times \\ 
&\times \,
\big[ \psi_1(kx) \psi_2(ky) - \psi_1(ky) \psi_2(kx) \big],
\\
\Ftilde_x^{\mu\rho} \Ftilde^*_{y \rho\nu}
&=
\phantom{-} \frac{m^2}{e^2} \xi_1\xi_2 \, k^\mu \Lambda_{5\nu}
\,\times \\ 
&\times \,
\big[ \psi_1(kx) \psi_2(ky) - \psi_1(ky) \psi_2(kx) \big].
\end{aligned}
\end{gather}

Using these relations we can show that for $j=j'=0$
\begin{align}
\uplambda^\mu 
&= 
- 2m\tau \sum_{i=1,2} \Lambda_i^\mu \xi_i I_i,
\nonumber\displaybreak[0]\\\nonumber
\Ftilde^{\mu\nu}_x \uplambda_\nu
&=
\phantom{-}2 \frac{m^2}{e} \tau k^\mu \sum_{i=1,2} \xi_i^2 \psi_i(kx) I_i,
\nonumber\displaybreak[0]\\
\Ftilde^{*\mu\nu}_x \uplambda_\nu
&=
-2 \frac{m^2}{e} \tau \Lambda_5^\mu \xi_1 \xi_2
\big[ \psi_1(kx) I_2 - \psi_2(kx) I_1 \big],
\end{align}
\begin{align}
\Lambda_i\uplambda
&=
2m \tau \xi_i I_i,
\nonumber\displaybreak[0]\\\nonumber
e q\Ftilde_{x}\uplambda 
&=
2 kq \,\tau  m^2 \sum_{i=1,2} \xi^2_i \psi_i(kx) I_i,
\nonumber\displaybreak[0]\\\nonumber
\Lambda_{i\mu}^* k_{\nu} \eps^{\mu\nu\rho\sigma} q_\rho \uplambda_\sigma &= 
2m\tau kq \, \xi_i I_i,
\nonumber\displaybreak[0]\\\nonumber
\Lambda^*_{1\mu} \Lambda^*_{2\nu} \eps^{\mu\nu\rho\sigma} q_\rho \uplambda_\sigma 
&= 0,
\nonumber\displaybreak[0]\\\nonumber
\Lambda_{1\mu} k_{\nu} \eps^{\mu\nu\rho\sigma} q_\rho \uplambda_\sigma &= 
\phantom{-} 2m\tau \xi_2 I_2 \, q\Lambda_5,
\nonumber\displaybreak[0]\\\nonumber
\Lambda_{2\mu} k_{\nu} \eps^{\mu\nu\rho\sigma} q_\rho \uplambda_\sigma &= 
- 2m\tau \xi_1 I_1 \, q\Lambda_5,
\nonumber\displaybreak[0]\\
\Lambda_{1\mu} \Lambda_{2\nu} \eps^{\mu\nu\rho\sigma} q_\rho \uplambda_\sigma &= \phantom{-} 0
\end{align}
and
\begin{gather}
e\Lambda_{i\mu} \Ftilde^{\mu\nu}_x q_\nu
= 
m\, kq\, \xi_i\, \psi_i(kx).
\end{gather}

\end{document}